\documentclass[12pt]{article}
\usepackage[utf8]{inputenc}
\usepackage{hyperref}
\usepackage{textcomp, gensymb}
\usepackage[utf8]{inputenc}
\usepackage{amsmath}
\usepackage{amssymb}
\usepackage{xcolor}
\usepackage{float}
\usepackage[numbers]{natbib}
\usepackage{csquotes}
\usepackage{caption}
\usepackage{url}
\usepackage{cite}
\usepackage[a4paper, margin = 1in]{geometry}
\usepackage{graphicx}

\title{Systematization of Knowledge: Synthetic Assets, Derivatives, and On-Chain Portfolio Management}

\author{Abrar Rahman\footnote{Corresponding author contact: abrarfrahman@berkeley.edu}  \qquad Victor Shi \qquad Matthew Ding \qquad Elliot Choi \medskip \medskip\\ University of California, Berkeley}

\date{September 20, 2022}

\begin{document}
\maketitle
\begin{abstract}
    Synthetic assets are decentralized finance (DeFi) analogues of derivatives in the traditional finance (TradFi) world—financial arrangements which derive value from and are directly pegged to fluctuations in the value of an underlying asset (ex: futures and options). Synthetic assets occupy a unique niche, serving to facilitate currency exchange, giving traders a means to speculate on the value of crypto assets without directly holding them, and powering more complex financial tools such as yield optimizers and portfolio management suites. Unfortunately, the academic literature on this topic is highly disparate and struggles to keep up with rapid changes in the space. We present the first Systematization of Knowledge (SoK) in this area, focusing on presenting the key mechanisms, protocols, and issues in an accessible fashion to highlight risks for participants as well as areas of research interest. This paper takes a broad perspective in establishing a general framework for synthetic assets, from the ideological origins of crypto to legal barriers for firms in this space, encapsulating the basic mechanisms underpinning derivatives markets as well as presenting data-driven analyses of major protocols.
\end{abstract}

\newpage
\tableofcontents
\newpage

\section{Overview}
\subsection{Goals of the Standardization of Knowledge (SoK)}
\begin{enumerate}
    \item Understand the key components, principles and models of the protocols
    \begin{itemize}
        \item For each protocol category, describe the financial problem that it solves.
        \item What are the key performance indicators (KPIs) to evaluate the capabilities of the protocols?
    \end{itemize}
    
    \item Relevant design mechanisms for the protocols
    \begin{itemize}
        \item Derivatives pegging/generation mechanisms
        \item Price divergence reduction
        \item Market liquidity guarantees
        \item Risk management strategies
        \item Decentralized clearance mechanisms
    \end{itemize}
    
    \item New protocol designs \& future directions for the space
\end{enumerate}

\subsection{Methods to Organize the Derivatives Space}
A comparative analysis of various decentralized finance protocols ought to keep the following characteristics in mind to meaningfully evaluate the effectiveness of the system design for all key stakeholders [\citeonline{nyfed}]:

\begin{enumerate}
    \item Trading Risk: If a user wants to sell the derivative, what is the mechanism for doing so? Is there enough liquidity available on the market to trade effectively?
    
    \item Counterparty Risk:
    \begin{itemize}
        \item How are margins determined? Ex: collateral ratios, types of collateral assets.
        
        \item How is counterparty risk mitigated? What are the costs and benefits of participating in the market to the various stakeholders? Ex: liquidation protocols.
    \end{itemize}
    
    \item Tracking Error (Price Divergence):  What algorithms ensure that the price of a given derivative asset tracks to that of the reference asset to minimize capital loss to arbitrage opportunities [\citeonline{arbitrage_pdiv}]? 
    
    \item Identifying Natural Buyers: For each protocol, are natural buyers and sellers clearly defined? Derivatives perform a critical economic function for these stakeholders because they allow hedgers and speculators to mutually trade a specific risk. Do holders of a given synthetic asset have any real utility from doing so, or is the market primarily controlled by speculators? 
\end{enumerate}

\section{Introduction}
\subsection{Ideological and Philosophical Perspectives}

Investors, developers, and policymakers alike should take great efforts to understand the ideological underpinnings of the cryptocurrency revolution before diving into the ins-and-outs of the status quo of the space. Blockchain is a foundational technology, not an application-driven one. It aims to fundamentally change incentive structures, which are ultimately a narrow, specific set of socially-motivated concerns with respect to digital autonomy and financial privacy, as opposed to being a catch-all silver bullet for the evolution of the global financial system. Ultimately, all technologies have their trade-offs, and decentralized finance protocols are no different. The explosion of new tokens, exchanges, and technologies—and the brazen speculation thereof—has fuelled the rise of blockchain to the top of the current zeitgeist. But with speculatory fervor comes inherent risks. Thus, analysts must reckon with fundamental questions about the intent and application of decentralization in all its forms in order to gain a deeper, holistic understanding of the space. 

The key question one ought to ask before crafting or investigating a new token-based architecture is \emph{Why Decentralize}? In order for any crypto product to pass the smell test, it must have a compelling answer for this underlying concern. As a new foundational architecture, blockchain has challenged conventional assumptions about the mechanics of where global finance is and how it could or ought to operate. In doing so, most applications of the technology are quite experimental and should be regarded as such for the foreseeable future. Even the most carefully crafted smart contracts rely on key assumptions about user behavior, which may or may not be validated. And if a virtual asset is pumped by speculators in lieu of gaining meaningful adoption by natural users, no amount of past data or theory will give a definitive answer on the practical utility or stability of its market.

One can look to the early adopters of cryptocurrencies for insight into the soul of the technology. Though crypto is relatively new, the ‘cypherpunk’ subculture that spawned it goes back to at least 1994, with the publication of ``The Crypto Anarchist Manifesto" [\citeonline{crypto_manifesto}]. The following excerpt is near-required reading for anyone interested in the motivations for the crypto revolution:

\begin{displayquote}
Computer technology is on the verge of providing the ability for individuals and groups to communicate and interact with each other in a totally anonymous manner. Two persons may exchange messages, conduct business, and negotiate electronic contracts without ever knowing the True Name, or legal identity, of the other. Interactions over networks will be untraceable, via extensive re-routing of encrypted packets and tamper-proof boxes which implement cryptographic protocols with nearly perfect assurance against any tampering...Combined with emerging information markets, crypto anarchy will create a liquid market for any and all material which can be put into words and pictures. And just as a seemingly minor invention like barbed wire made possible the fencing-off of vast ranches and farms, thus altering forever the concepts of land and property rights in the frontier West, so too will the seemingly minor discovery out of an arcane branch of mathematics come to be the wire clippers which dismantle the barbed wire around intellectual property. Arise, you have nothing to lose but your barbed wire fences!
\end{displayquote}

The manifesto is eerily prescient for a quarter-century old piece of writing. Taking cues from anarchist thought, this letter established the crypto-anarchist milieu, which seeks to preserve privacy rights while ensuring digital citizens are empowered to act and associate on their own volition [\citeonline{chohan}]. These concepts became the common ideology of what went on to be known as the cypherpunk forums of the late 1990s and early 2000s.

The still-unnamed inventor (going by ``Satoshi Nakamoto") of the first cryptocurrency was inspired by ethical concerns from the cypherpunk forums. The blockchain data structure, first devised solely to develop Bitcoin [\citeonline{nakamoto}], has grown to become the foundation of decentralized finance. The blockchain is an append-only decentralized public ledger serving as an secure, immutable clearing house. Paraphrasing a friend of the authors, Blockchain @ Berkeley President Darya Kaviani, the blockchain is a technological solution to work around the social problem of corruption in centralized institutions [\citeonline{bab}]. 

While Bitcoin popularized decentralized computing, it is limited by design as a straightforward currency. Ethereum pushed the envelope by encoding “smart contracts,” enabling more complex capital flows which have become the basis for decentralized finance protocols and decentralized autonomous organizations (DAOs) [\citeonline{vitalik}].

Businesses built on smart contracts can automate away much of the human discretion and biases baked into decision-making in traditional firms, which can be applied selectively to hardcode secure, ethical business practices [\citeonline{sulkowski}], algorithmically guaranteeing compliance with financial regulations. However, policymakers should be wary of novel sociotechnical approaches that create unprecedented ethical buffer zones. Protocols become baked-in, and people are either stripped of agency or rendered unnecessary. Smart contract-driven business operations lack many of the human checks and balances that keep our global financial system stable. Thus, regulators, entrepreneurs, and laypeople alike must decide how much human agents can simply give way to algorithmically controlled enterprise.

\subsection{Synthetic Assets \& Crypto Derivatives}

Synthetic assets are essentially tokenized derivatives. In the traditional financial (TradFi) world, derivatives are positions on stocks and bonds that a trader does not own [\citeonline{derivative_def}]. If a trader seeks to profit from speculating on the price fluctuations of a stock that they do not own, they can do so through derivatives markets. Synthetic assets have taken this process one step further by immutably tracking fluctuations of the original asset on the blockchain, creating a cryptocurrency token to represent the price movement of the underlying asset [\citeonline{crypto_deriv_def}]. Synthetic asset derivatives have naturally gained traction within the decentralized finance (DeFi) realm, seeing as they allow for seamless hybrid portfolios that hedge against crypto volatility by incorporating TradFi options [\citeonline{deloitte_hedge}].

Using a derivative to tie the value to an already existing asset and then create a token for this derivative, investors can theoretically trade \emph{anything} on Web3 architectures. One of the main reasons why synthetic assets are emerging as an increasingly preferred method of investing is because of the added security and traceability [\citeonline{hbr_truth}]. While traditional trading happens on centralized exchanges (ex: the New York Stock Exchange), with synthetic assets, all trades are securely, immutably, and decentrally stored on the blockchain, all while granting traders the freedom of optional anonymity. \\
  
\noindent To summarize, the advantages of asset tokenization are [\citeonline{mirror_white}]:
\begin{enumerate}
    \item \textbf{Reduced Geographical Barriers} - It is easy to access all the information and records of previous transactions since they are all stored on a permissionless blockchain, and individuals can transact from anywhere in the world.
    \item \textbf{Reduced Reliance on Middlemen} - In TradFi, there needs to be a trusted middleman to validate and facilitate transactions, but it is eliminated thanks to blockchain's immutability and transparency. 
    \item \textbf{Enhanced Accessibility through Fractional Ownership} - Like stocks, tokenized assets can be divided into as many units (tokens) as desired, thereby enabling wider investment participation for high-value assets such as real estate and expensive stocks. White fractional ownership for stocks is becoming increasingly popular in brokerages, it comes with operational overhead that does not exist with tokenized stocks.
    \item \textbf{Improved Asset Liquidity} - Valuable assets that are hard to trade/transfer tend to suffer in terms of liquidity when exiting. Blockchain substantially reduces friction and therefore permits higher liquidity. 
    \item \textbf{Increased Transaction Efficiency} - Compared to traditional settlements, blockchain transactions can dramatically improve efficiency by reducing time and cost. 
    \item \textbf{Expanded Investor Base} - Fractional ownership and permssionless blockchain improve access to investment opportunities by allowing investors to partake in transactions that were previously inaccessible to them due to capital or liquidity constraints.
\end{enumerate}

\subsubsection{Outline of a Typical Derivatives Protocol}
\noindent The following process outlines the process commonly used to create synthetic assets:
\begin{enumerate}
    \item Choose a real-world asset (ex: USD, ETH, TSLA, Gold, etc.) to underlie the synthetic derivative asset.
    \item Leverage oracles: data feeds that query the off-chain underlying asset data while applying measures to reduce tracking error [\citeonline{chainlink_faq}].
    \item Create (``mint") synthetic asset and start trading!
\end{enumerate}

\subsection{Stablecoins}
Though not derivatives in the strictest sense of the term, stablecoins are a popular choice for DeFi traders, serving to: hedge against volatility in standard tokens, facilitate forex/derivatives trading with synthetic currencies, or as the underlying tokenomic mechanism underpinning a decentralized liquidity pool or yield optimizing protocol. There are four possible ways to collateralize a stablecoin [\citeonline{centre_whitepaper}]:
\begin{enumerate}
    \item Fiat-Collateralized: pegging tokens to reserves of fiat currencies;
    \item Crypto-Collateralized: maintaining reserves of valued tokens like BTC or ETH;
    \item Algorithmically Non-Collateralized: software models aim to provide tokenomic price stability without relying on underlying collateralized assets;
    \item Hybrid: incorporating aspects of two or more of the above approaches.
\end{enumerate}

Due to the inextricable interconnectedness of DeFi markets, understanding the principles and theoretical limits of these coins is a prerequisite for a full appreciation of the dynamics of the derivatives space.

\subsubsection{USDC: Fully-Backed Virtual Assets} \label{usdc}
The USD Coin is a collaborative effort between Centre and Coinbase which is designed to maintain a one-to-one peg with the US Dollar [\citeonline{usdc}]. Centre has pushed for fiat collateral. This means that every USDC in circulation is backed by fiat assets in reserves, thus providing a surefire guarantee of price stability by pegging token value to reserved fiat value. The USDC token is available across most blockchains, in Ethereum ERC-20, Algorand ASA, Avalanche ERC-20, Flow FT,  Hedera SDK, Solana SPL, Stellar asset, and TRON TRC-20 forms.

\subsubsection{Central Bank Digital Currencies}
A central bank distributed currency (CBDC) is defined as a digital liability of a central bank that is widely available to the general public, a digital alternative to fiat currency [\citeonline{fedres_cbdc}]. Today, physical bank notes are the only type of central bank money available to the general public, while other financial institutions must leverage massive deposits of assets and purchase federal deposit insurance to enable digital payments via bank balances. A CBDC sacrifices the purported benefits of decentralization for the trustworthiness of no associated credit or liquidity risk, all while enabling interoperability with crypto wallet systems. The Federal Reserve defines the key characteristics of a CBDC as such [\citeonline{fedres_cbdc}]:
\begin{itemize}
    \item Privacy Protected: strike a balance between limiting criminal activity and ensuring consumer privacy
    \item Intermediated: the private sector, not the central bank, manages accounts or digital wallets to facilitate the management of CBDC holdings and payments
    \item Transferable: seamless integration between intermediaries to make national payments systems more efficient
    \item Identity-Verified: the most anti-crypto notion possible, but required for compliance with financial regulations
\end{itemize}

\noindent The proposed benefits include the following [\citeonline{fedres_cbdc}]: 
\begin{itemize}
    \item Safely Meet Future Needs and Demands for Payment Services
    \item Improvements to Cross-Border Payments
    \item Financial Inclusion
    \item Extend Public Access to Safe Central Bank Money
\end{itemize}

\noindent The Federal Reserve identifies the following risks and policy considerations [\citeonline{fedres_cbdc}]: 
\begin{itemize}
    \item Changes to Financial-Sector Market Structure
    \item Safety and Stability of the Financial System
    \item Efficacy of Monetary Policy Implementation
    \item Privacy and Data Protection and the Prevention of Financial Crimes
    \item Operational Resilience and Cybersecurity
\end{itemize}

Since 2014, over 60 central banks have experimented with CBDCs, and 14\% of central banks have projects at advanced development stages [\citeonline{pwc_cbdc}]. CBDCs are further subdivided into two major categories: retail coins, which are intended for use by the general public as a full, equally-supported alternative to physical banknotes; and interbank or wholesale coins, which are accessible only to banks to settle transactions on financial assets, intended instead to accelerate the integration of distributed ledger technologies into existing market infrastructures. 70\% of declared wholesale projects are already running pilots, while only 23\% of retail projects have reached this implementation stage.

\subsubsection{Algorithmic Stablecoins}
While fiat-collateralized tokens like USDC and some implementations of CBDCs have strong pegging mechanisms in place, which are shared to a lesser extent by crypto-collateralized coins, there is much more debate over whether algorithmically non-collateralized coins can ever truly guarantee stability long-term. Algorithm-based stablecoins seek to use algorithms built on economic models to increase or decrease the supply of stablecoins in response to changes in demand [\citeonline{bis_stablecoin}]. 

The use of stablecoins has increased in terms of transactions (a measure of capital flow) to parallel the growth in market capitalisation (a measure of overall assets locked) [\citeonline{bis_stablecoin}]. While stablecoins are designed to be less vulnerable to speculative bubbles, their market capitalisation still fluctuates greatly with buys and sells by investors. Additionally, many fear that without additional private or public sector circuit-breakers in place, stablecoins that crash may trigger sell-offs in any assets that they are pegged to, creating broader market chaos, raising interest in non-collateralized approaches. However, without any underlying asset to justify the value of the coin, these sell-offs often prove to be more pronounced and possibly unrecoverable for algorithmic stablecoins.

The design of public algorithmic stablecoins as based on two mechanisms: 1) the collateralized mechanism and 2) the algorithmic peg mechanism [\citeonline{fedres_stablecoin}]. For example, Dai tokens are minted when a user deposits a standard volatility cryptocurrency (ex: ETH). In return, the user is loaned Dai, which is designed to be pegged to USD, at a greater than 100\% collateralization ratio. The loan automatically liquidates if the value of the collateralization token drops below a given threshold. Dai actually uses a hybrid collateral model which includes fiat-backed stablecoins (ex: USDC). The algorithmic peg mechanism, in contrast, uses automated smart contracts to defend the by buying and selling the stablecoin against the governance token. These algorithmic pegging systems are inherently vulnerable. They rely on incentivizing independent actors on the market to perform price-stabilizing arbitrage, which requires reliable price information at all times [\citeonline{calgary_stablecoin}]. During periods of high volatility, this house of cards often inevitably falls apart.

Algorithmic stablecoins are vulnerable to two main sources of volatility where the underlying economic assumptions conflict with the observed market dynamics [\citeonline{stablecoin_volatility}]. The first is broken expansionary phases. The mechanism of expanding the number of tokens on the market serves to increase the supply of of a stablecoin when it rises over the one dollar peg due to increases in demand. Periods of unexpectedly high demand can lead to a extremely high yield rate, further fuelling a bubble of speculatory buying. In a similar vein, the opposite can occur in broken contraction. The market for a stablecoin may experience low participation in contraction, i.e., many investors are unwilling to burn the staking token for the stablecoin due to the fear that they might never be able to redeem, a lack of faith in the underlying digital asset that may prove fatal to the assumptions of any algorithmic stablecoin's tokenomic design.

\subsubsection{Terra: the So-Called Stablecoin} \label{terra}
The inextricable interconnectedness of stablecoin systems with the derivatives space is better explained by example as opposed to by theory, as individual implementations of different tokenomic models vary greatly in how they leverage stablecoins.

As a case study in what can potentially go awry in a derivatives market with a supposedly stable algorithmic token, examine the Terra blockchain, a supposed architecture for multi-chain compatible stablecoins. Terra took a subtly, but meaningfully distinct approach to the design and bootstrapping of its network [\citeonline{terra_luna}]. For some background, consider the role that stablecoins play on the Ethereum network. The rise of  Ethereum-based ERC-20 stablecoins has improved the utility of the network by providing less-volatile alternatives to ETH, but bear in mind that Ethereum has no native stablecoin: the market has instead adapted by building extensions on top of it. In stark contrast, Terra used its native staking token (LUNA) to collateralize a set of algorithmic stablecoins, most notably UST, which was intended to be pegged one-to-one with the US dollar by maintaining a steady burn rate of LUNA via arbitrage opportunity incentivization. 

By designing the network with stablecoins in mind from the get-go, Terra was claimed to better leverage inflation rewards in the native token than other blockchains. Terra offered an absurd 19.5\% yield on staking—equivalent to earning 19.5\% interest on a loan—through the Anchor protocol [\citeonline{cnet_ust}]. This promise of high yields attracted many investors, but any model that relies on infinite yield from underlying asset movement should have raised skepticism. The problem is that this model assumes that the market for the staking token will remain stable, which is never guaranteed, and this vulnerability lead to the stunning collapse of the UST token.  

UST lost its dollar peg on May 9th, 2022 [\citeonline{verge_ust}]. Over \$2 billion worth of UST was unstaked from the Anchor protocol, of which hundreds of millions were immediately sold. It is still not determined whether this was an innocent selloff based off the current bull market for both crypto assets and the stock market, or if this was an intentional attack on the Terra network [\citeonline{cnet_ust}]. Regardless, this triggered unprecedented hyperinflation in the LUNA token, crashing its price from \$100 to less than 1 cent. Terra took the extraordinary step of shutting down their blockchain for a few days. Binance, which is the largest DEX in terms of trading volume, halted the trading of UST and LUNA in response. The OKX exchange swiftly followed. Meanwhile, as of May 14th, the FTX exchange among other DEXs have continued to list both tokens. The wipeout of over \$15 billion in market capitalization serves as a cautionary tale for algorithmic stablecoins in general, and may inspire regulatory backlash from governments worldwide.

\subsection{Potential Risks for Crypto Derivatives}
\subsubsection{Market Instability}

One of the chief motivators behind crypto derivatives trading is to speculate on the fluctuations in the prices of crypto assets. This model relies on the widely-held understanding that many token economies have historically been subject to instability due to the prevalence of speculatory day-trading, as opposed to long-term holdings or using tokens as meaningful currencies with predictable values. 

An analysis of Bitcoin by Two Sigma applied their traditional finance model, Factor Lens, and showed a largely idiosyncratic risk profile that was hard to predict, but had limited positive correlations with the global equity market [\citeonline{two_sigma}]. Notably, BTC generally behaved like an inflation sensitive asset. They also found that the 10 largest coins by volume were all positively correlated, with DOGE showing the least association. Positive associations between crypto assets suggests that a wallet with many coins might not reap significant diversification benefits.

In April 2022, the International Monetary Fund (IMF) released a comprehensive report on the volatility of virtual assets [\citeonline{imf}], which now have a significant impact on global capital flows. Central banks must shift their strategies to promote stability. Interestingly, as cryptocurrencies are often bought in USD, there has been a resulting increase in the dollarization of other global currencies. Additionally, though many investors cultivate a diverse portfolio of token offerings, there is high correlation between the prices of most coins as they tend to have high overlap in ownership and are not tied to market fundamentals. In a cruel twist of fate, cryptocurrencies may actually be increasing centralization for monetary and fiscal policymakers. These concerns have broad implications for the development and regulatory prospects of crypto derivatives markets.

\subsubsection{Differences Between TradFi and DeFi Regulatory Protections}

DeFi markets do not share the same protections that their centralized counterparts offer. TradFi firms are kept under a close watch by regulators. For example, banks pay dues to the Federal Deposit Insurance Corporation (FDIC), which backs deposits as a guard rail against bank runs [\citeonline{fdic_hist}]. The FDIC was established by 1933 Banking Act (Glass-Steagall), enacted in the middle of the Great Depression, during which fully a third of American banks failed [\citeonline{glass}]. Many DeFi protocols substitute the backing of the FDIC with smart contract based agreements with liquidity providers. 

There is no intermediary with the legal purview to monitor markets for fraud and manipulation, prevent money laundering, safeguard deposited funds, ensure counterparty performance, or make customers whole when processes fail. DeFi markets, platforms, and websites are not registered as DCMs or Swap Execution Facilities (SEFs). SEFs are mandated by federal law ever since the passage of the 2010 Dodd-Frank Act, a comprehensive set of Wall Street reforms passed in the aftermath of the Great Recession [\citeonline{cftc_dodd}]. The CEA does not exempt crypto-based services from having to register under existing legal frameworks designed to protect traders.

Many decentralized exchanges (DEXs) have implemented automated emergency stop measures, but without the legal/regulatory infrastructure in place, investors may remain wary about the liability of trading on grey-market platforms.

\subsubsection{Regulatory Changes in the Crypto Space}
Technology tends to advance one step ahead of legislators, and the DeFi space is no different. The White House recently issued the first Executive Order on blockchain technology, in an attempt to unify federal agencies' approaches to decentralized finance [\citeonline{biden}]. Though a grand unified approach to regulation is a key first step, current political instability means that this order is anything but set in stone. For example, consider how a hypothetical President Ron DeSantis might change the approach to digital asset regulation. Federal regulators have not yet "future-proofed" their policy directives. 

Wyoming is the first subnational entity that has taken initiative on creating new legal structures to accommodate DeFi enterprises. Their DAO LLC is the first decentralized autonomous organization legal entity with limited liability for stakeholders [\citeonline{wyoming}]. It remains to be seen how liability can be litigated in a business entity with no true leadership, but with smart contracts, the code is now the first evidence examined in a court of law, an unprecedented approach in the history of law.

On the federal level, some crypto finance platforms are now being regulated as securities. This opens up a vast world of existing legislation and case law, potentially opening the floodgates to litigation against Web3 financial service firms. The Securities and Exchange Commission (SEC) has taken increasingly aggressive action to protect investors consumers in this space. On May 3, 2022, the SEC nearly doubled the size of its Crypto Assets and Cyber Unit, with around 50 dedicated positions [\citeonline{sec_may2022_double}]. Since its creation in 2017, the unit has brought more than 80 enforcement actions related to fraudulent and unregistered crypto asset offerings and platforms, resulting in monetary relief totaling more than \$2 billion. 

Here is a non-exhaustive list of pertinent regulatory actions undertaken by the SEC in recent years, and why they are notable:
\begin{itemize}
    \item (8/6/2021) The first regulatory action against any DeFi firm was a fraud case [\citeonline{sec_aug2021_bcp}]. Gregory Keough, Derek Acree, and their company Blockchain Credit Partners were found to have sold securities in unregistered offerings through DeFi Money Market from February 2020 to February 2021, and additionally lied to investors about holdings in car loans to back their securities.
    \item (2/14/2022) BlockFi was the first DeFi company to come to a financial compliance agreement with the Securities and Exchange Commission (SEC) [\citeonline{sec_feb2022_blockfi}]. SEC Chair Gary Gensler said, ``Today’s settlement makes clear that crypto markets must comply with time-tested securities laws, such as the Securities Act of 1933 and the Investment Company Act of 1940. It further demonstrates the Commission’s willingness to work with crypto platforms to determine how they can come into compliance with those laws." 
    \item (5/6/2022) The SEC charged NVIDIA Corporation, one of the leading Graphics Processing Unit (GPU) firms, with inadequate disclosures about the impact of cryptomining to its bottom line [\citeonline{sec_may2022_nvidia}]. GPUs are the preferred kind of chip for mining due to their enhanced parallel computing capabilities. In two of its Forms 10-Q in 2018, NVIDIA reported the growth in its GPU sales within its gaming business. NVIDIA had known that this increase in gaming sales was driven in significant part by cryptomining, and declined to properly disclose to investors the true source of its revenue gains.
    \item (5/6/2022) The SEC halted Mining Capital Coin from operating their fraudulent crypto enterprise [\citeonline{sec_may2022_mcc}]. Before investors' one-year ``memberships" expired, they encountered purported errors that stymied their efforts and were required to either buy another mining package or forfeit their investments. “As the complaint alleges, [the founders] took every opportunity to extract more money from unsuspecting investors on false promises of outlandish returns and used investor funds raised from this fraudulent scheme to fund a lavish lifestyle, including purchasing Lamborghinis, yachts, and real estate,” said A. Kristina Littman, Chief of the SEC Enforcement Division’s Crypto Assets and Cyber Unit.
\end{itemize}

\subsubsection{Legality of Crypto Derivatives}

Most significantly, DeFi markets for derivatives (ex: token-based futures contracts) may not be legal under the Commodity Exchange Act (CEA), a U.S. law that governs such products and requires them to trade only on regulated designated contract markets (DCMs), according to Commodity Futures Trading Commission (CFTC) Commissioner Dan Berkovitz in a speech to the Asset Management Derivatives Forum [\citeonline{cftc_ban}].

Due to the potential regulation risk, Uniswap protocol removed the trading interface on its front-end for some synthetic derivatives [\citeonline{uniswap_nosynth}], though the back-end protocol still supports them. Uniswap is one of the largest DEXs, which hit an all-time high transaction volume in Q4 2021 with \$238.4 billion  worth of trades, more than 61\% above Q3 2021 [\citeonline{messari_2021q4}].  Interestingly, this action was taken less than a month after Berkovitz's speech [\citeonline{uniswap_nosynth}], and the blacklisting was mostly restricted to assets at risk of being classified as securities, including tokenized stocks, options tokens, insurance-based tokens and synthetic assets from crypto derivatives platforms like Synthetix.

\section{Related Works}
This paper takes a broad, Systematization-of-Knowledge (SoK) approach to documenting the crypto derivatives space. Though there have been a number of meta-analyses of decentralized finance protocols, prior works have different focus areas that do not fully encompass the broad scope of this paper. 

\subsection{Automatic Market Makers}
An SoK last revised on January 2022 focusing on Automated Market Makers (AMMs) establishes a general framework describing the tokenonomics of the state-space representation for AMMs and systematically compares the top AMM protocols' mechanics [\citeonline{amm_sok}]. They further discuss security and privacy concerns, emerging from the underlying properties and assumptions of AMM-based DEXs. 

Synthetic derivative markets tend to utilize AMM-based DEXs, so these perspectives are pertinent for this papers' scope. However, their main area of analysis is restricted to the markets themselves, as opposed to the on-chain assets.

\subsection{L2 Bridges}
Another SoK released in December 2021 focuses instead on Layer 2 bridges between various blockchains [\citeonline{bridge_sok}]. As on-chain transactions on any cryptographically-secure distributed ledger technology (DLT) system are inherently computationally expensive, the scalability of blockchain-based architectures has been a leading concern for stakeholders in the space. 

Thus, there has been substantial work done to find methods to a) move computationally intensive processes off-chain, b) create secure validation bridges to seamlessly integrate off-chain transactions, and c) leverage L2 bridges to efficiently integrate transactions across multiple blockchains. 

Bridges are a key technology that allows synthetic asset markets to provide liquidity pools access across multiple chains. This paper takes a more cybersecurity-focused approach to the space than is strictly useful for a coherent understanding of the dynamics of synthetic asset derivatives.

\section{Major Protocols}

\subsection{Synthetic Assets}

\subsubsection{Synthetix}

Synthetix is a derivatives liquidity protocol allowing anyone, to gain on-chain exposure to a vast range of assets [\citeonline{synthetix}]. It enables the creation of synthetic assets called Synths, which are derivative tokens tracking the price of other assets, including cryptocurrencies, foreign exchange markets (forex trading), indices, equities, and commodities. Synths are backed by the SNX native token, which is staked as collateral, entitling stakers to fees generated by Synth trades.

Synthetix allows people to gain exposure to assets on Ethereum they might not otherwise be able to access. It also solves problems of liquidity and slippage, as all trades between Synths are operated on peer-to-peer contracts. The protocol does all this without compromising on decentralization. Traders retain custody of their funds at all times. Other decentralized trading options like AMM’s have major liquidity and slippage restrictions for most assets. Synthetix also allows traders to borrow synthetic assets against ETH and begin trading immediately rather than needing to sell ETH.\\

\noindent \emph{Key Mechanisms:}
\begin{enumerate}
    \item Collateral: Stake SNX as collateral to generate sUSD for the Synths trading in the entire system. The collateral ratio is 400\% on L1 and 750\% on L2. Mintr is the staking platform for SNX [\citeonline{mintr}]
    \item Neutral Debt Pool: Use for pooling all the synthetic assets in the investment for all the users, and manage the risk and revenue. 
    \item Unlimited Liquidity: Synthetix uses high trading fees to avoid the arbitrage for the delay of the oracle under the mechanism of unlimited liquidity.
\end{enumerate}

\noindent \emph{Ecosystems on L2:}
\begin{enumerate}
    \item Kwenta: trading platform for all Synths (including futures) on L2 [\citeonline{kwenta}]
    \item Lyra: options on L2 [\citeonline{lyra}]
    \item Thales: prediction markets and binary options on L2 [\citeonline{thales_why}]
\end{enumerate}

\noindent In order to minimize deviations from the peg (collateralization ratio), Synthetix offers three solutions, or \emph{synthetic pegging mechanisms} [\citeonline{synthetix}]:
\begin{enumerate}
    \item Arbitrage: SNX stakers create debt by minting synthetic assets. If the peg drops, stakers can profit from buying sUSD back below par and burning it to reduce their debt.
    \item sETH liquidity pool on Uniswap: a portion of SNX added to the total supply (due to inflationary policy) is distributed as a reward to sETH/ETH liquidity providers.
    \item SNX auction: partnering with dFusion protocol, discount SNXs are sold for ETH, which is then used to purchase synthetic assets below peg.
\end{enumerate}

\subsubsection{Mirror}
Mirror is a DeFi protocol powered by smart contracts on the Terra network that allows for the creation of ``Mirrored Assets” (mAssets), which track with the price of real world assets [\citeonline{mirror_docs}]. Mirror is an interchain protocol, with cross-compatibility with Ethereum and the Binance Smart Chain (BSC). Unlike the Synthetix protocol, the Mirror Protocol uses an automated market maker (AMM) for trading mAssets. 

Mirror Protocol’s native token \$MIR is distributed as a reward for providing mAssets liquidity to AMMs. This acts as an operational subsidy for having to calibrate the collateralization ratio. When the value of the asset rises above the collateralization threshold, the collateral is liquidated for the solvency of the system. \$MIR also acts as a governance token for stakers to have a say in the policies of the platform.

In order to successfully mint an mAsset, a synethic version of a real-world asset, there are four main operations [\citeonline{mirror_docs}]: 
\begin{itemize}
    \item \textbf{Mint:} Collateral is needed to mint an mAsset. The collateral could be a stablecoin or a different mAsset. However, the required minimum ratio is different for different collateral (150\% for stablecoin collateral, 200\% for mAsset), and the ratio is agreed upon by DAO governance. 
    \item \textbf{Burn:} When the trader wants to exit, the mAsset is burned, and the corresponding amount of collateral is returned. 
    \item \textbf{Trade:} mAssets are traded on an AMM on any compatible blockchain. 
\end{itemize}

Correspondingly, there are four main types of participants, each with different roles and incentives[\citeonline{mirror_docs}]:

\begin{itemize}
    \item \textbf{Minter:} A minter is a user who provide the necessary collateral and mint mAssets. Essentially, a minter is taking a short position on the assets, so the minter is being the risk of price changes. When the price of the asset raise, the minter is forced to deposit more to maintain the minimum ratio or get liquidated. Conversely, if the price decreases, the minter can take some collateral out. The minter's counterparty is the system itself.
    \item \textbf{Trader:} A trader is a user who buy and sell mAssets on decentralized exchanges supported by Mirror and benefits from price exposure via mAssets. 
    \item \textbf{Liquidity Provider:} A liquidity provider provides necessary liquidity to the market by adding equal amounts of an mAsset and UST to the liquidity pool. In return, the liquidity provider gets trading fees from the pool and also liquidity provider (LP) tokens, which represents the provier's share in the pool. To exit, LP tokens can be burned to redeem corresponding amount of mAssets and UST.  
    \item \textbf{Staker:} Staker can deposit LP tokens or MIR tokens to earn rewards.  
\end{itemize}

When a minter is not able to maintain the minimum collateral ratio, liquidation is triggered. Mirror protocol would seize a portion of the collateral, and the mAsset would be sold at a discount. The process continues until the collateral ratio is above the minimum ratio. 

Terra historically has used inflation of their tokens to incentivize early adopters to host their networks. As network effects proliferate with growth, slowly the reliance on inflation of the underlying virtual asset decreases, but adoption and utility of the network ought to gradually replace the mining/staking revenue stream with transaction fees. 

However, being a Terra ecosystem, the health of LUNA and UST is very crucial for the performance of Mirror. As UST can no longer maintain its peg and sinks below \$1, it takes far more UST tokens to buy one of the synthetic assets, since UST prices decrease while the prices of the underlying assets remain the same. In other words, it costs considerably more to buy these assets, or it now earns considerably less from these assets. Users and investors are existing the protocol, and MIR market cap has fallen from 90 million to only 20 million. See Section \ref{terra} for more on the UST meltdown.

\subsubsection{Anchor Protocol}
Anchor Protocol is a savings protocol on the Terra blockchain that seeks to find a middle-ground between highly volatile crypto assets and the more conventional low-yield stablecoin offerings like Compound [\citeonline{anchor_whitepaper}]. It is built on block rewards of major Proof-of-Stake blockchains. Anchor promises a principal-protected stablecoin savings financial product. 

The core building block of the Anchor savings protocol is the Terra money market, a Web Assemby smart contract that facilitates depositing and borrowing of Terra stablecoins (TerraUSD, for instance). This money market is defined by a pool of Terra deposits that earn interest from borrowers. Borrowers put down digital assets as collateral to borrow Terra from the pool. The interest rate is determined programatically as a function of supply and demand from borrowers, which is inferred by the pool’s \emph{utilization ratio} $u$, where $u(t)$ is the fraction of Terra $(t)$ in the pool that has been borrowed:

$$u(t) = \frac{deposits_{borrowed}}{deposits_{total}}$$

\noindent \emph{Algorithmic Determination of Interest Rates:}
\begin{itemize}
    \item Anchor uses block rewards across blockchains to derive DeFi’s benchmark rate. With Anchor, the return that depositors can expect is a function of borrowers’ on-chain income. 
    \item The Anchor money market is a unique enabler of “yield transfer” from borrower to depositor by accepting bAssets as collateral. The resulting diversified yield, the Anchor Rate, reflects the market’s preferred sources of yield on the blockchain. Anchor markets this as "The Gold Standard for Passive Income on the Blockchain".
    \item The Anchor Rate is the average of the rolling yields of all bAssets used as collateral for borrowing the stablecoin, weighted by the aggregate (Terra-denominated) collateral value of each asset.
    \item The Anchor smart contract dynamically distributes block rewards from collateral bAssets between borrower and depositor to achieve the target rate.
    \item Loan to Value (LTV) ratios range from 0 to 1 and are a function of an asset’s volatility and liquidity. Stable, liquid assets will have high LTV ratios, while volatile illiquid assets will have low LTV ratios.
\end{itemize}

The Anchor liquidation protocol claims to ensure safe maintenance of deposits safety by paying off debts that are at risk of violating collateral requirements.
\begin{itemize}
    \item To ensure that all Anchor loans are sufficiently collateralized, the liquidation protocol pays back “at risk” loans using “liquidation contracts” undertakes the task of paying back debt in exchange for collateral plus a fee – the “liquidation fee”. Liquidation contracts can be written by anyone, are aggregated in a pool and tapped “on demand” when a loan needs to be liquidated in ascending order of liquidation fee. In addition to the liquidation fee, contracts earn a passive premium charged to borrowers that is calibrated to ensure full coverage of outstanding loans.
    \item The structure and incentives built into liquidation contracts enable them to provide higher robustness and solvency guarantees compared to a traditional “keeper” system. Keeper systems rely on arbitrageurs to finance liquidations on a discretionary basis, which can result in a liquidity crunch at times of high market volatility. 
    \item This risk has materialized in practice in Maker’s keeper system, resulting in huge losses for borrowers. Liquidation contracts, on the contrary, are fully collateralized and enforce a lengthy withdrawal period. Liquidation demand is therefore predictable and stable in the face of temporary shocks, thus protecting both depositors and borrowers.
\end{itemize}

The Anchor Protocol notably failed to prevent the meltdown in the Terra Network, covered in depth in Section \ref{terra}.

\subsection{Oracles}

\subsubsection{Chainlink}
Chainlink is a decentralized oracle network which uses a reputation system to establish trusted data sources for real-world assets to empower crypto derivative assets [\citeonline{chainlink_whitepaper}]. Smart contract applications rely on data feeds and APIs that are external to the blockchain. The operations and distributed governance of blockchain architectures mean that they cannot directly fetch data in real-time. Oracles prior to Chainlink were primarily centralized, fulfilling a standard middleman role instead of leveraging the power of decentralized, trustless systems. 

Their whitepaper describes both the on-chain smart contracts to connectivity components, as well as the off-chain software software powering the nodes of the network. Consensus mechanisms are done off-chain consensus. Designed with aggressive adversarial data feed attackers in mind, the system has reputation and security monitoring services so synthetic asset protocols can develop derivatives from reliable data sources that have achieved network consensus. 

Chainlink node operators receive compensation in the form of LINK tokens. Nodes are responsible for: 1) retrieval of data from off-chain data feeds, 2) reformatting data into blockchain-friendly formats, and 3) off-chain computation. To bridge across multiple chains, the node operator determines the exchange rate from the native token to LINK based on demand for data on the off-chain resource their oracle provides. Other oracle operators provide competition, ensuring that rates are fair to all stakeholders. LINK is an ERC-20 token with the additional ERC-223 “transfer and call” functionality, allowing tokens to be efficiently received and processed by contracts within one transaction.

\subsubsection{UMA}
UMA, which stands for Universal Market Access, is an oracle built to prioritize the efficient deployment of data feeds, a core technology enabling many synthetic asset protocols to scale [\citeonline{uma}]. UMA’s Optimistic Oracle allows contracts to quickly request and receive various types of data. The Optimistic Oracle acts as a generalized escalation game between contracts that initiate a request and UMA’s dispute resolution system known as the Data Verification Mechanism (DVM).

Data proposed by the Optimistic Oracle will not be sent to the DVM unless it is disputed. This enables contracts to obtain information within any predefined length of time without the need to have the data recorded on-chain.

The UMA Long Short Pair (LSP) contract template allows for the creation of non-liquidatable and capped payout products via tokenized long and short positions. Developers can choose between Binary, Linear, Covered Call, and Range LSP contract templates.

Call and put options allow DAO treasuries to offer yield opportunities to tokenholders, as well as leverage opportunities to speculators. The options are traded against the project token, so DAOs can create liquidity without any opportunity cost.

Range tokens offer an alternative way for DAOs to diversify their treasury without selling their native tokens. DAOs can use their native tokens as collateral to borrow stablecoins with zero risk of liquidation. These collateralization incentive structures can be built with requirements for certain scores on project-specific key performance indicators (KPIs) that are governed by smart contracts. Smart airdrops motivate the community to create content, proselytize relentlessly, increase adoption and establish corporations.

\subsection{Derivatives Trading}

\subsubsection{dYdX}

dYdX is a DEX that takes a hybrid, off-chain order book approach based on the 0x API [\citeonline{dydx_whitepaper}]. It allows market makers to sign and transmit orders on an off-blockchain platform, in which the blockchain infrastructure is reserved exclusively for settlement. Financial products can be traded at any price agreed upon two parties, meaning there is no requirement for underlying contracts to be aware of the market price. There are three [\citeonline{dydx_prices}] pricing mechanisms available on dYdX:
\begin{itemize}
    \item Mid Market Price: the average between the lowest ask (sell price) and highest bid (buy price) on the 0x orderbook
    \item Index Price: aggregate based off data from multiple exchanges, used to trigger stop orders. In line with much of dYdX's design philosophy, it is managed off-chain to minimize delay and slippage in stop order triggering. The multi-exchange aggregation protects traders from flash crashes on any single exchange
    \item Oracle Price: an aggregate price calculated using multiple on-chain price oracles, used for collateralization and liquidations
\end{itemize}

\noindent There are three main smart contracts that underlie this DEX.
\begin{itemize}
    \item Proxy: used to transfer user funds
    \item Margin: offers functionality to enable margin trading as it contains all business logic and public functions
    \item Vault: holds all funds locked up in position
\end{itemize}

A trader opens a margin position by sending a transaction to the dYdX margin smart contract containing a loan offer, a buy order for the borrowed token, and the amount to borrow. Upon receiving this transaction, the smart contract transfers the margin deposit from the trader to itself, and then uses an external decentralized exchange (DEX) such as 0x to sell the loaned token using the specified buy order.	The smart contract holds onto the deposit and token resulting from the sale of the loaned token for the life of the position. The position is closed when the trader sends a transaction to the smart contract containing a sell order offering to sell the amount of token owed to the lender for an amount less than or equal to the amount locked in the position. 

Upon receiving this transaction, the contract uses an external DEX to execute the trade between the order maker and itself. Afterwards, the contract sends the owed amount of the loaned token to the lender.	The amount owed to the lender includes the interest fee. The trader is sent all of the leftover token, which is equal to deposit + profit. Note that the profit could very well be negative if the price moved against the position!

The margin trading protocol can be used for both short selling and leveraged long trading by simply switching which token is borrowed (referred to as the owed token) with the one that is held in the position (referred to as the held token). The protocol allows the margin deposit to be paid in either token. If the deposit is paid in the owed token, it is sold along with the owed token borrowed from the lender, so that only the held token is, well, held in the position. 

Similarly, the payout to the trader from closing can be in either token. If the payout is in the owed token, all held tokens in the position are sold for the owed token and whatever is leftover after paying the lender is paid out to the trader.

When used for short selling, the trader will borrow a base token from the lender, which will be sold for a quote token. The trader puts up a margin deposit in the quote token. Only the quote token is actually held in the position. When the position is closed, the base token will be bought and paid back to the lender, and the trader will be paid out in quote token.

When used to take a leveraged long position, the trader will borrow the quote token from the lender, and put up a margin deposit in quote token. The quote tokens (some borrowed from the lender, and some put up as margin deposit) are then sold for the base token. Only the base token is held in the position. When the position is closed all of the base token is sold for quote token. The quote token is paid back to the lender, and the trader is again paid out in quote token. \\

\noindent \emph{dYdX Insurance Fund Mechanics}

Offering perpetual markets with a high amount of leverage inevitably entails increased risk. In particular, during times of high volatility in the underlying spot markets (tied to each perpetual contract), it is possible that the value of some accounts will drop below zero before they can be liquidated. Should these “underwater” accounts occur, they must be handled promptly in order to ensure the solvency of the system as a whole.

The insurance fund is the first backstop to maintain the solvency of the system when an account has a negative balance. The account will be liquidated, and the insurance fund will take on the loss. This insurance fund is the first emergency measure applied before any deleveraging occurs, with the following constraints:

\begin{itemize}
    \item The initial seed amount for the fund was 600,000 USD
    \item The insurance fund account and its activities are publicly auditable and verifiable
    \item The insurance fund off-chain. The dYdX team will be directly responsible for deposits to and withdrawals from the fund
    \item In the future, dYdX may decentralize some aspects of the fund; but the current priority is to ensure that underwater accounts are dealt with in a timely manner
\end{itemize}

In the event that the insurance fund is depleted, positions with the most profit and leverage may be used to offset negative-balance accounts, in order to maintain the stability of the system. Deleveraging is a feature made available by the perpetual contract, which is used as a last resort to close underwater positions if the insurance fund is depleted. Deleveraging works similarly to “auto-deleveraging” in other high-leverage futures and perpetual markets, and is a mechanism which requires profitable traders to contribute part of their profits to offset underwater accounts. This system is reserved with the following constraints:
\begin{itemize}
    \item Deleveraging will only be used if the insurance fund is depleted.
    \item Deleveraging is performed by automatically reducing the positions of some traders, prioritizing accounts with a combination of high profit and high leverage, and using their profits to offset underwater accounts.
    \item Deleveraging is chosen over a socialized loss mechanism to reduce the uncertainty faced by traders trading at lower risk levels.
    \item The most highly leveraged offsetting accounts will be deleveraged onto first.
\end{itemize}

\noindent \emph{dYdX Deleveraging Example:}
\begin{itemize}
    \item Assume an initial margin requirement of 10\% and a maintenance margin requirement of 7.5\%.
    \item Trader A deposits 1000 USDC, then opens a long position of 1 BTC at a price of 2000 USDC. Their account balance is -1000 USDC, +1 BTC. During a period of intense and prolonged volatility, the index price reaches 1080 USDC. Trader A is in a risky position, but not yet liquidatable. The price then rapidly drops further, and before A can be liquidated, the index price reaches 900 USDC, making the nominal value of A’s account -100 USDC.
    \item The insurance fund is already depleted due to recent price swings, so deleveraging kicks in. Trader B, whose current balance is 10000 USDC, -9 BTC, is selected as the counterparty, on the basis of B’s profit and leverage, and the fact that B’s short position can offset A’s long position.
    \item Trader B receives A’s entire balance, leaving A with zero balance, and bringing B’s total balance to 9000 USDC, -8 BTC. Trader B’s nominal loss due to deleveraging is 100 USDC, at an index price of 900 USDC. Trader B’s margin percentage increased (and leverage decreased) as a result of deleveraging, from 23.46\% to 25\%.
\end{itemize}

\noindent Scaling Solutions:
\begin{itemize}
    \item Margin: L1
    \item Perputual: L2 StarkEX (ZK-Rollup, Validium, and Volition)
\end{itemize}

\subsubsection{Deri Protocol}

Trades are executed through an AMM and positions are tokenized into non-fungible tokens (NFTs), which are composable with other projects [\citeonline{deri}]. It claims to be built following the given principles:

\begin{itemize}
    \item Real DeFi: Deri Protocol is a group of smart contracts deployed on the Ethereum blockchain, where the exchange of risk exposures takes place completely on-chain.
    \item Real derivative: The Profit and Loss (PnLs) of the users’ positions are calculated with mark price updated by oracle, which ensures precision; positions are maintained by a margin, which provides built-in leverage.
    \item Composability: Positions are tokenized as NFTs, which can be held, transferred or imported into any other DeFi projects for their own financial purposes (as blocks in their own “lego game”).
    \item Openness: anybody can launch a pool with any base token (but usually with a stablecoin, ex: DAI). The protocol does not enforce any preferred base token.
    \item Simplicity: Deri protocol adopts an extremely simple trading process.
    \item Intercompatibility: \$Deri Token is intercompatible and supports three different Blockchains through the Cross-Chain Deri Bridge (Ethereum, Binance Smart Chain, and Huobi Eco Chain)
\end{itemize}

Instead of having expiration dates like standard options, users must pay a fee to maintain their options. The funding fee is set to be proportional to the spread of the mark price divided by the index price. Every second, a long position maintains a payment to a short position funding fee as below:
$$F = f*(P-i)$$
$P$ is the mark price, $f$ is the funding fee coefficient, and $i$ is the index price. Bear in mind that the funding fee $F$ is positive when the long is higher than the short position, so in this case long positions pay short positions, and vice versa when $F$ is negative.

With Deri's Proactive Market Making (DPMM) algorithm, when the net difference between the long and short positions is 0, that means that the mark price equals the index price fed by the oracle. Whenever there is a trade, it pushes the mark price toward a trading direction (selling pushes down, buying pushes up). The price change due to the trade is proportional to the trade size. The price spread and the mark price are determined by the total net position, as below:
$$\Delta P / i = a(l-s)$$
$l$ and $s$ are the long and short positions, $\Delta P$ is the price spread, and the coefficient $a$ is particular to the pool liquidity and the pool parameters.

Since the v2 update, three additional features have been introduced to significantly increase capital efficiency to an extreme level:
\begin{itemize}
    \item Dynamic mixed margin: Accepts traders to choose one or more from the supported base tokens to post as margin.
    \item Dynamic liquidity providing: Accepts liquidity providers to choose one or more from the supported base tokens to provide as liquidity.
    \item Multiple trading symbols in one pool: Supports multiple trading symbols within one single pool
\end{itemize}

\subsubsection{Lyra}
Lyra is an options and futures L2 trading protocol within the Synthetix ecosystem. Its pool structure accepts stablecoins as collateral and offers options in rounds, which are 28 day periods with options tradable with four discrete expiry dates (at 7, 14, 21, and 28 days, respectively). Liquidity is split between two sub-pools: The Collateral Pool serves to, well, collateralize options and pays/receives premiums. The Delta Pool hedges delta exposure [\citeonline{deltaexp}] by trading the underlying asset for cash (as opposed to a futures market). Delta adjusted exposure is a measure that describes the first order price sensitivity of an option (or a set of options in a portfolio) to changes in the price of an underlying security. For an option of value $V$ and underlying asset price $S$:
\begin{equation}
dV \approx \Delta S \frac{dS}{S} = \Delta _{\$} \frac{dS}{S}
\end{equation}

There are a number of automatic hedging protocols in place to mitigate risk for holders. The first is simply market-driven pricing: options are priced using market based, strike-adjusted IVs applied to the Black-Scholes-Merton model, returning a theoretical value (W). The second is Vega risk management. The protocol charges with an asymmetric spread around W based on whether a given trade increases or decreases the LP’s Vega risk. Finally, Delta Risk Management, in which the protocol hedges the liquidity pool’s net delta by trading the underlying asset on a spot market. After factoring in the flat fee (f) for trading on any given exchange, the trade executes with a final transaction cost of W + f.

Scaling solution: Optimistic Rollup

\subsubsection{Vega}  
Vega is an end-to-end margin trading protocol secured on proof-of-stake [\citeonline{vega_whitepaper}]. Notably, it is blockchain-agnostic and has no fees on orders, and incurs no gas fees on the Ethereum blockchain. Through a system of token governance, anyone can create a derivative market on any underlying asset and build liquidity by automatically matching markets to traders and market makers. Liquidity providers of these community-curated markets can earn incentives on every trade, creating a venture capital like approach encouraging incubation of a portfolio of markets.

Vega maintains a blockchain agnostic approach by decoupling financial products from the underlying smart contracts by means of an in-house ``Smart Product" language, allowing for a black-box approach where complex financial products can work on any chain with any underlying asset. Smart Products only directly interface the protocol when settling a position, and allocates collateral intelligently through a chosen risk model.\\

\noindent Markets operate in one of six modes:
\begin{enumerate}
    \item Continuous Trading: the standard mode of operation.
    \item Discrete Trading: Trading via frequent batch auctions.
    \item Auctions: Used during market creation, upon the resumption of trading after suspension due to illiquidity, or resolving the price during extreme price moves.
    \item Suspended Market: Suspension is only used when the market has broken down. It operates akin to an auction with a call period with no defined end: orders will be accepted, but no trades will be executed.
    \item Request for Quote (RFQ): Intended for more illiquid assets. Traders and market makers are free to make prices at their discretion. Once a quote is accepted, a trade will result and be settled like any other transaction.
    \item Matched Trades: Any two participants can submit the same or compatible trade details within a given time window, and the Vega network will automatically accept and manage the transaction like an RFQ driven trade.
\end{enumerate}

Any network user who wishes to be a market maker can become one in a permissionless fashion, and there is no restriction on the minimum amount of liquidity one needs to provide in order to qualify. There are two subcategories of market makers. Active market making requires participants to manage a pricing strategy for their market making volume, and they will be penalized if not updating their pricing stratification regularly. One can also passively make a market, supplying liquidity algorithmically using an average of the risk portfolios of other market makers, though passive market makers get a lower reward for their limited participation.

Rewards for market makers, the liquidity providers for Vega, are calculated in proportion to the total volume that is traded on the market, and the percentage of the reward that an individual market maker receives depends on the following:
\begin{itemize}
    \item their market making stake (relative to the total)
    \item their price making activity (active and competitive prices receive greater rewards)
    \item the longevity of their market making commitment
\end{itemize}

Each community market has an associated insurance pool. It serves as collateral protection
in case of a shortfall. The insurance pool starts with no funding on the creation of the market, but will earn funds over time from both trading activity and released insurance funds upon the maturation of other markets. On each transaction, the stochastic risk model determines allocations of fees to or payments from the insurance pool depending on the directionality of slippage. The insurance pool will also gain funds from penalties when a trader attempts a double-spend or if a market maker fails to meet their liquidity obligations. When a market closes, either due to a predetermined order-book expiration date or a governance vote, the insurance pool is divided between other markets that share the same base currency as its underlying collateral asset.\\

\noindent The whitepaper highlights the following risks for participants on the network:
\begin{itemize}
    \item Credit Risk: seeing as participants are only identified by their public key, there is no recourse if a trader owes more in settlement than their posted collateral. Thus, it is critical that the protocol design constantly maintains significant collateralization for all positions.
    \item Liquidity Risk: Liquidity is a key assumption for fair trades in all markets. If either the buy or sell side of the order book volume does not have sufficient liquidity to close the largest single counter position held by a trader, the market is suspended, and will be reopened with an auction-first system conditional on the provisioning of additional liquidity by market makers or traders.
    \item Extreme Price Moves: In the event of a drastic price move, the protocol includes a `circuit breaker' which switches the market from a continuous/discrete open trading model to an auction system, for fairer price discovery.
    \item Risks from Decentralization: controlling a sufficient stake within a node network gives an attacker total control of the consensus algorithm, which grants them control over all the funds in margin accounts and insurance pools.
\end{itemize}
  
\subsubsection{Thales}
 
Thales [\citeonline{thales_why}] is a binary option trading protocol within the Synthetix ecosystem, which is effectively an on-chain permissionless betting market. Via a series of Chainlink-provided price feeds, users can hedge and speculate on the prices of synthetic assets, commodities, sports markets, and derived performance metrics for all of the aforementioned categories. 

Options smart contract payouts depend strictly on whether predetermined condition is met or not. The sSHORT option token is a bet the condition will not be met, whereas the sLONG option token is a bet the condition will be met. On the expiry date, if the condition is met, the long option tokens are worth 1 sUSD and the short option tokens are worth 0 sUSD; or vice versa. In the rare case in which the underlying condition finishes exactly at the strike price on the strike date, the smart contract logic supports resolving in favor of long options for cases where the underlying price is strictly equal to the strike price to two decimal points! 

Peer-to-peer trading occurs by placing orders on 0x order books [\citeonline{0x_orders}]. 0x is an API serving as the infrastructure for a  DEX that enables the exchange of Ethereum-type assets on multiple chains.

Anyone with can create a betting market on Thales with the following procedure: 
\begin{itemize}
    \item Choose an asset to create a market for
    \item Input a strike price for the chosen asset
    \item Choose and input a market maturity date and time
    \item Input the amount of sUSD to fund the market with (min. 1000 sUSD)
\end{itemize}	

For liquidity provision, a straightforward minting model [\citeonline{thales_model}] is used in which anyone can mint options by depositing sUSD. For every 1 sUSD, a user receive 0.99 sLONG and 0.99 sSHORT (1\% minting fee; fee is shared between protocol fee pool and market creator). For example, if Abrar deposits 1000 sUSD, he receives 990 sLONG and 990 sSHORT binary options tokens. 
 
Scaling solution: Optimistic Rollup
 
\subsection{Yield Optimizers}

\subsubsection{Yearn Finance} 

Yearn Finance is a suite of products in Decentralized Finance (DeFi) that is designed to generate yield on smart contract platforms like Ethereum [\citeonline{yearn}]. The protocol is maintained by various independent developers and is governed by YFI holders. The system runs on BTC and ETH.

Capital pools that automatically generate yield based on opportunities present in the market. Vaults benefit users by socializing gas costs, automating the yield generation and rebalancing process, and automatically shifting capital as opportunities arise. End users also do not need to have a proficient knowledge of the underlying protocols involved or DeFi, thus the Vaults represent a passive-investing strategy.

After depositing, your funds first go to the vault contract and then are deployed to one or more strategy contracts. Guardians and strategists monitor deposits in order ensure optimal returns and to be available during critical situations.

The first Yearn product was a lending aggregator. Funds are shifted between dYdX, AAVE, and Compound automatically as interest rates change between these protocols. Users can deposit to these lending aggregator smart contracts via the Earn page. This product completely optimizes the interest accrual process for end-users to ensure they are obtaining the highest interest rates at all times among the platforms specified above.\\

\noindent \emph{v2 yVault Improvements}
\begin{itemize}
    \item Up to 20 strategies per vault: This will increase the flexibility to manage capital efficiently during different market scenarios. Each strategy has a capital cap. This is useful to avoid over allocating funds to a strategy which cannot increase APY anymore.
    \item Strategist and Guardian are the new Controllers: The Controller concept is not available in V2 yVaults and has been replaced by a Guardian and the Strategy creator (strategist). These 2 actors oversee strategy performance and are empowered to take action to improve capital management or act on critical situations.
    \item Automated vault housekeeping (Keep3r network): harvest() and earn() calls are now automated through the Keep3r bots network. These 2 function calls are used to purchase new underlying collateral by selling the earned tokens while moving the profits back to the vault and later into strategies. The keep3r network takes the heavy lifting of doing these calls and running with the gas costs in exchange for keep3r tokens. This approach unloads humans from these housekeeping tasks.
    \item Bouncers and Guest lists: Yearn has created an unique development process for new vaults. All vaults are launched as Test Vaults (tyvToken) to start with. Test vaults have a cap and therefore their strategies as well. Also, the Bouncer has a guest list of wallets which can interact by depositing and withdrawing funds in the Test Vaults. This approach prevents uninformed users from potentially losing funds in a not production ready product.
\end{itemize}

\subsubsection{Abracadabra Money}
Abracadabra Money (Abra) is a decentralized crypto lending platform that allows users to collateralize digital assets in exchange for Magic Internet Money (MIM) stablecoin loans. However, differing from most lending platforms, it takes interest-bearing tokens (ibTKNs), which is a type of liquidity provider (LP) token. [\citeonline{abracadabra_shrimpy}] In order to attract liquidity, more and more DeFi protcols, such as Yearn Finance and Sushi, reward liquidity providers with LP tokens. Those tokens can increase in value and generate returns, and Abra allows users to borrow against those tokens and further maximize returns. 

Abra has an isolated lending market mechanism. In most Defi lending platforms, if you deposit several types of collateral, your risk across each is the same. But Abra uses Kashi, an isolated risk market that allows you to open different collateral debt positions (CDPs) for each asset you deposit. Abra is also engineered for cross-chain compatibility, as you can take your MIM across different chains 

To ensure liquidity and the stability of its stablecoin MIM, Abra is giving the Curve holders SPELL, which is Abra’s governance token, to help maintain Abra’s stablecoin 
Abracadabra is passing a vote to permanently funnel 5\% of all revenues to bribe Convex for these votes, forever securing the appeal of the MIM stablecoin pool to ensure it stays pegged to \$1. In addition, Abracadabra partnered with Olympus DAO to take control of its liquidity. Protocol-owned liquidity, or POL means the deep liquidity it rents is under its permanent discretion. POL allows protocols like Abracadabra to focus on delivering better products without having to worry about its liquidity seeping away.

The Olympus Pro program enables that security through a process called Bonding. With Bonding, LPs sell their liquidity to Abracadabra in exchange for discounted SPELL tokens. That gives LPs an instant profit on their position paid in SPELL and mitigates the risks of impermanent loss. At that point, they can cash out or stake the SPELL for additional gains. Abra also offers different farming opportunities where users can stake their LP tokens to farm SPELL tokens. This mechanism is used to keep deep liquidity on particular pairs. Farming is currently available on Ethereum for ETH-SPELL and MIM-3CRV LP tokens. Future incentivization programs for LP pairs can be voted on through the governance process when the governance portal goes live. [\citeonline{abra_doc}]

Incentivizing liquidity providers with SPELL tokens worked well to get the word out on Abracadabra and achieve deep multichain liquidity. But issuing SPELL rewards dilutes the SPELL token and keeps the platform locked in battle against other DeFi upstarts with better rewards.

\subsubsection{Harvest Finance}

Harvest Finance is a yield farming protocol that lets users put their assets to work in high-producing farming opportunities. Notably, Harvest does not provide services to consumers in the US for fear of regulatory backlash, as it does not conform to TradFi financial disclosures [\citeonline{harvest}]. The system uses BTC and ETH.

As Decentralized Finance (DeFi) has grown, it has become increasingly common practice for protocols to provide token rewards in return for utilizing their platform, generally in the form of `staking' tokens. Collecting these rewards is generally referred to as `yield farming' or 'liquidity mining.' Harvest Finance acts as an intermediary in this process, collecting tokens from a large pool of stakeholders and staking them en-masse. The interest bearing token for the protocol is called iFARM, which works by being staked in the profit sharing pool.

On a regular basis, Harvest Finance automatically collects or `harvests' the reward tokens and exchanges them for more of the underlying assets that users deposited, thereby compounding the interest they earn. This is claimed to save users in terms both time and the network fees associated with making multiple transactions.

Investment strategies are tied to particular vaults through a `timelock' mechanism. When a change is first announced, a 12 hour period passes before the change is made official and taken live. All strategies are called via doHardWork() to compound their interest with the following steps:
\begin{enumerate}
    \item Collect reward tokens accumulated by the invested funds.
    \item Liquidate the rewards into the invested token.
    \item Invest recovered funds into the investment opportunity.
\end{enumerate}

Whenever doHardWork() is called, the price of each share of the associated vault increases. Most vaults are harvested every 12-48 hours, depending on the liquidity of their associated token. During liquidation, as of September 8th, 2021, the DAO governance structure has decided that 70\% of the yield farming revenue is returned to users who provide capital, while the remaining 30\% of the yield farming revenue is distributed to users who stake FARM in the Profit Sharing contract.

\subsubsection{Autofarm Network}

Autofarm is a one-stop DeFi suite comprising of 6 main products (Autofarm - yield optimizer/vaults, AutoSwap DEX Aggregator, AutoPortfolio, AutoTrade, AutoAnalytics) [\citeonline{autofarm}]. The Autofarm ecosystem's biggest advantage is its compatibility with 15 EVM chains: BNB Chain (Formerly Binance Smart Chain) (BNB), Polygon Network (MATIC), Huobi ECO Chain (HECO), Avalance chain (AVAX), Fantom chain (FTM), Moonriver chain (MOVR), OEC Chain (OKB) (OKEx), xDai Chain (xDai), Cronos Chain (CRO), Boba Chain (BOBA), Aurora (NEAR), Celo Chain (CELO), Harmony (ONE), Velas (VLX), and Oasis (ROSE).

Otherwise, this is a straightforward yield aggregator following the same design principles of most competitors. The specifics of the Autofarm business model are as follows:
\begin{itemize}
    \item Controller fee: 0.4\% on profits to controller
    \item Platform fee: 0.5\% on profits to platform
    \item Entrance fee: < 0.1\% on capital to pool
    \item Withdrawal fee: none
    \item Fee to AUTO staking vault: 2.5\% on profits
    \item Total cost: < 0.1\% on capital to pool + 3.4\% on profits
\end{itemize}

\subsection{Portfolio Management}

The primary indicator used to compare the adoption of these protocols is assets under management (AUM), serving a similar function to total value locked (TVL) in general DeFi protocols.

\subsubsection{Coinbase}
Notably \emph{not} an on-chain portfolio management tool, Coinbase is still one of the largest players in the DeFi space and thus a platform to which every other player is compared. 

Coinbase offers a plethora of products in its ambition to be the primary trading interface for cryptocurrencies and DeFi tokens [\citeonline{coinbase_prod}]. For individual investors, it offers a trading portal, wallet, credit card, lending solutions, and even NFTs. For businesses and developers, it has a suite of more focused asset hosting, brokerage, analytics, cloud, and oracle tools. It is also one of the main forces behind USDC, explored in-depth in section \ref{usdc}. They even have a venture capital division, which invests in other DeFi companies.

Coinbase is a major force in DeFi which has even advertised at the Super Bowl. Coinbase was the first crypto company to go public, at a valuation of over \$100 billion. As of December 2021, the company had over \$21.2 billion in assets [\citeonline{coinbase_sec}]. However, Coinbase wallets are notably custodied by the exchange, as opposed to the user. This means that Coinbase can easily blacklist users on their centralized exchange, and in the event of bankrupcy, may seize assets from users to deleverage their overall holdings. This is a major shortcoming that has kept up interest in decentralized alternatives.

\subsubsection{TokenSets}

Set Protocol is an Ethereum-native DeFi primitive that leverages existing Open Finance protocols to allow for the bundling of crypto-assets into fully collateralised baskets, which are represented as ERC-20 tokens on the Ethereum blockchain [\citeonline{sets_docs}]. These Set tokens act as structured products that represent the manager's strategy, which others can replicate by simply holding the Set. 

The underlying contract that enables management of the Set supports external integrations with exchanges, lending platforms, automated market makers, and asset protocols, also enabling more advanced strategies employing not only DEX trades but yield farming and margin trading.

There are three roles allocated for in this protocol:
\begin{itemize}
    \item Investor: allocate capital into performance-proven products
    \item Manager (Set creator): create a structured product to offer to investors, and earn the streaming fee
    \item Developer: build a bespoke application on top of Set
\end{itemize}

\noindent \emph{Key concepts:}
\begin{itemize}
    \item Set Creator - view the Set creators wallet via Etherscan.
    \item Market Cap - derived from the total assets under management within the underlying Set contract deployed on Ethereum
    \item Streaming Fee - the method a set is monetized by the Manager that is payable by the Investor over time.
    \item Performance - everyone wants to track the performance of the assets they invest in and currently Zerion tracks the performance of all created Sets on the protocol out of the box for the entire duration. It's worth noting that if the Set has enough liquidity on a DEX or is listed on a CEX it would also by default be chartable on the likes of TradingView and other 3rd party charting services.
    \item Current Balance - the total dollar value of the total amount of that particular Set token you own. This is based on the Wallet you are connected with to TokenSets.
    \item Allocation Table - The Allocation Table is where you will see all the key information regarding what assets are backing each Set token! It's important to remember that Sets are 100\% collateralised baskets of Tokens. This means that the value of a Set is directly related to the underlying assets and will never be any discrepancy on the matter. In the above example, the DeFi Pulse Index is a sector-based structured product with a central aim to capture the performance of DeFi based projects using a detailed methodology.
\end{itemize}

\noindent Here are the claimed advantages of this architecture:
\begin{itemize}
    \item No Impermanent Loss - Uses market capitalization for component weights, not fixed percent, so it does not suffer underperformance from impermanent loss.
    \item Cheap \& Efficient - Eliminating the need to perform countless costly transactions manually saves you time and money.
    \item Diversified Portfolio - trends less volatile than more concentrated portfolios. Downside protection due to holding a wider selection of tokens.
\end{itemize}

The business model relies on a streaming fee, which is the method by which a set is monetized by the Manager that is payable by the Investor over time. It is an annualized fee on the total market cap of the set that the Set Creator may accrue at any time. For example, a Set is created with a 2\% streaming fee. After 6 months, the Set Creator may claim 1\% of the total market cap of the Set as fees.

The DeFi Pulse Index is a digital asset index designed to track tokens’ performance within the Decentralized Finance industry. The index is weighted based on the value of each token’s circulating supply. The DeFi Pulse Index aims to track projects in Decentralized Finance that have significant usage and show a commitment to ongoing maintenance and development. As part of the inclusion assessment, DeFi Pulse considers a wide range of characteristics of the token, the project and the protocol. These criteria can be placed in four groups:
\begin{enumerate}
    \item Token’s Descriptive Characteristics
    \begin{enumerate}
        \item The token must be available on the Ethereum blockchain.
        \item The token must be associated with a decentralized finance protocol or dapp listed on DeFi Pulse.
        \item The token must not be considered a security by the corresponding authorities across different jurisdictions.
        \item The token must be a bearer instrument. None of the following will be included in the index:
        \begin{enumerate}
            \item Wrapped tokens
            \item Tokenized derivatives
            \item Synthetic assets
            \item Tokens that are tied to physical assets
            \item Tokens that represent claims on other tokens.
        \end{enumerate}
    \end{enumerate}
    \item Token’s Supply Characteristics
    \begin{enumerate}
        \item It must be possible to reasonably predict the token’s supply over the next five years.
        \item At least 7.5\% of the five year supply must be currently circulating.
        \item Token must have reasonable and consistent DeFi liquidity on Ethereum.
        \item The token’s economics must not have locking, minting or other patterns that would significantly disadvantage passive holders.
    \end{enumerate}
    \item Project’s Traction Characteristics
    \begin{enumerate}
        \item The project must be widely considered to be building a useful protocol or product.
        \item Projects focused on competitive trading/holding, having Ponzi characteristics, or projects that exist primarily for entertainment, will not be included.
        \item The project’s protocol must have significant usage. 
        \item The protocol or product must have been launched at least 180 days before being able to qualify to be included in the index.
        \item The protocol or project must not be insolvent.
    \end{enumerate}
    \item Protocol’s User Safety Characteristics
    \begin{enumerate}
        \item Security professionals must have reviewed the protocol to determine that security best practices have been followed to maintain user assets safe under different circumstances.
        \item Alternatively, the protocol must have been operating long enough to create a consensus about its safety in the decentralized finance community.
        \item In the event of a safety incident, the team must have responded promptly and addressed the incident responsibly in the aftermath, providing users of the protocol with a reliable solution and the decentralized finance community with adequate documentation to provide transparency about the incident. 
        \item The selected tokens must have sufficient liquidity across a variety of trading platforms.
    \end{enumerate}
\end{enumerate}

The Index Value is the spot value of the index. The Circulating Supply is the number of tokens circulating the last time circulating supply was determined. The first circulating supply was determined on September 8, 2020, using CoinGecko as a reference source. In this context, price is the market price of the token in USD. The Index Divisor is a constant that is adjusted on each rebalance. As of December, 2020, the Divisor was 35092014.36. The index is maintained monthly in two phases:
\begin{enumerate}
    \item Determination Phase
    \begin{enumerate}
        \item The determination phase takes place during the third week of the month. It is the phase when the changes needed for the next reconstitution are determined.
        \item Additions and deletions: The tokens being added and deleted from the index calculation are determined during the third week of the month and published before monthly reconstitution.
        \item Circulating Supply Determination: The DeFi Pulse Index currently references CoinGecko’s circulating supply number. The Circulating Supply is determined during the third week of the month and published before the monthly reconstitution.
        \item Weighting: Any token that has a weight greater than 25\% during the determination phase will have its weight capped at 25\%. Any excess weight will be redistributed to the remaining components of the DeFi Pulse Index on a weighted basis. This same process will be repeated for every token exceeding the 25\% allocation cap.
    \end{enumerate}
    \item Reconstitution Phase
    \begin{enumerate}
        \item Following publication of the determination phase outcome, the index composition will change to the new weights on the first working day of the following month. I.e components will be added or removed, and weights adjusted.
        \item Any funds based on the Defi Pulse Index will be expected to execute any buy and sell transactions on or shortly after the index reconstitution date.
    \end{enumerate}
\end{enumerate}

\subsubsection{Enzyme}
Enzyme is a decentralised asset management infrastructure built on Ethereum [\citeonline{enzyme}]. Using Enzyme Smart Vaults, individuals and communities can build, scale, and monetze investment (or execution) strategies.

Enzyme enables Depositors to interact with Vaults in a way which is non-custodial and requires minimal trust between parties. Vault Managers are held accountable by positive or negative measure, meaning they can be bound to performing certain actions or forbidden from others. Furthermore, live accounting is available for both performance and fees which is provable with on-chain data using the Enzyme subgraph. The subgraph also enables them to retrieve other useful reporting data (e.g. trade history or deposits/withdrawals) and make them transparently available to users. Policies such as fees or behaviors around investments can be quickly configured and imposed at the granularity of the Vault level.

At the moment, the protocol is transitioning towards an AUM based fee which is expected to be implemented in Q3 2021. The fee will be 25bps on Assets Under Management and will directly align interest between users of the network and the value of the MLN token, which is primarily tasked with incentivizing development, maintenance, and security of the network.

Organizations get a quick \& easy way to connect to multiple DEX trading venues (eg. DEXs, aggregators, derivatives), DeFi protocols (eg. depositing to AMM pools, lending) and DeFi benefits (eg. farming orairdrops) from one place. Enzyme provides users with full, easily readable, auditable, and transparent reporting over the AUM. Organizations often delegate trading to 3rd-party for efficiency in a non-custodial way (with the power to revoke said privilege at any time). Organizations can pull in outside investors or even bots to invest alongside their strategies and charge optional fees.

\subsubsection{PieDAO}

PieDAO is a community DAO focused on tokenising automated wealth creation strategies [\citeonline{piedao}]. The DAO operates using the following structure:
\begin{itemize}
    \item The DAO: The on-chain governance of PieDAO, managed via an Aragon DAO.
    \item Pie Smart Pools: The first DAO-governed Index funds managed as AMM pool.
    \item PieVaults: The first crypto index product with productive assets.
    \item The Oven: A system designed to democratize minting Pies by lowering costs.
\end{itemize}

The core products of PieDAO are carefully curated baskets of assets, traditionally known as ETFs — or Pies as they are referred to in this ecosystem. Pies are designed for people without much starting capital as a source of passive income given sufficient contribution to the governance structure. DOUGH, the PieDAO governance token, is the driving force behind the entire ecosystem, allowing people across the globe to make decisions on what products to create, what strategies to deploy, and what assets to include.

These Pies are designed to be gas-efficient on the ETH blockchain, using the following participation model:
\begin{enumerate}
    \item Users deposit their ETH and as soon as the threshold is reached the Oven automatically mints everyone’s pies.
    \item Withdraw your ``freshly baked pie" to your wallet.
    \item Oven checks if there’s enough ETH in the contract every fifteen minutes, automatically triggering minting if there’s over 10 ETH ready.
    \item Note: this process is obviously not completely gas-free as you do have to send the initial transaction and then withdraw, but by the DAO's estimation, the collective Oven protocol leverages their treasury balance to save users up to 97.5\% of the up-front cost.
    \item A baking transaction can need 73 transactions and cost as much as 4m GAS. This can be hugely expensive when GAS prices are high. With Oven, users just need a deposit and withdrawal, costing around 100k GAS.
    \item Currently, Oven can be used to mint DEFI++ (tneir diversified DEFI index) and BCP (Balanced Crypto Pie, containing equal parts BTC, ETH and DEFI++).
\end{enumerate}

PieDAO emphasizes distributing yield by active DAO participation. A Treasury Committee is being formed, whose main tasks will be ensuring sufficiently diversified allocation of assets, determining farming strategies to establish consistent returns to token holders, and bringing the DAO to self-sustainability. Any user can formally propose a new Pie through a proposal (PIP) on the forum. Once the governance system has reached a consensus on how to proceed, the following steps are executed:
\begin{enumerate}
    \item Computing the 30 Day Average Circulating Market Cap for each underlying asset.
    \item Considering the 30d Market Cap to weight in each asset allocation, applying an ``x” \% Cap to their allocation (exceeding allocation would be iteratively redistributed among other assets), getting to an Adjusted Market Weight
    \item Introducing a Sentiment Score to weight in for all underlying assets the shared sentiment according to the 3 following pillars
    \begin{enumerate}
        \item Innovation (Solution \& Roadmap/Timeline)
        \item Functionality to the DeFi Ecosystem (Uniqueness \& Composability)
        \item Growth Potential (P/E Ratio, Volume \& Outstanding Supply)
    \end{enumerate}
    \item Each of the above gets graded on a 1-to-12 scale, resulting in a Sentiment Score for each underlying asset.
    \item Calculate a weighted average with the Adjusted Market Weight (80\%), and the Sentiment Weight (20\%) derived from the Sentiment Score.
    \item Compute the number of tokens constituting the collateral for each asset as a function of its live price (Coingecko price feed), and its resulting Overall Weight allocation.
\end{enumerate}

\subsubsection{Hord}

Hord's model is to follow market leaders and communally stake assets following a few primary sources [\citeonline{hord}]. Influencers, asset managers, and any high-performing traders can create and update a public portfolio with:
\begin{enumerate}
    \item Initial Deposit
    \item Public Token Picks
    \item Weight of Each Asset
\end{enumerate}

Participants can join a real-time mirrored portfolio following a so-called Champion. Participants opt in a Hord smart contract that mirrors a Champion's portfolio, issuing an ETF token that reflects the Champion's portfolio assets movements and fluctuation in real-time. Champion followers can buy these ETF tokens that are minted when the follower deposits funds in the mirror smart contract. That ETF's value fluctuates in sync with the value of the assets in the Champions basket and can be redeemed or traded.

\subsubsection{DHedge}

dHEDGE is a non-custodial, permissionless, decentralized asset management tool for synthetic assets on the Ethereum blockchain powered by the Synthetix derivatives liquidity protocol. It relies on a system of ``managers" that, similar to Hord, are ranked based on their relative performance to offer example portfolios for individual investors to model their investment strategies off of. 

Managers make decisions on how to direct liquidity pool of funds, competing for dHEDGE leaderboard rankings. Based off of their performance, the manager may optionally extract a performance fee as a percentage of the overall return generated by the pool. Fees are collected in pool tokens. This means that the manager's incentives are to serve their pool dutifully, seeing as their profit is ultimately tied to increases in their overall ownership of the pool as well. Performance fees are notably not decided by any individual investor's profit, but instead, the manager's lifetime performance measured as a profit margin.

Managers invoke dHEDGE smart contracts to create pools. Managers may have public pools allowing anyone to become an investor, or create private pools that allows only certain Ethereum wallet addresses to become investors. Managers can use a range of methods to cultivate their pool's strategy, from active management or algorithmic approaches, to even investing in other pools on dHEDGE. Interestingly, managers can verify themselves on Twitter.\\

\noindent \emph{dHEDGE Leaderboard Ranking}
\begin{itemize}
    \item $Score = Sortino_{Ratio} * \sqrt{7\:Day\:Average\:Pool\:Value}$
    \item The Sortino Ratio is a measure of risk-adjusted returns. \begin{itemize}
        \item Standard Formula [\citeonline{sortino}], where $R_p$ is the actual portfolio return, $r_f$ is the risk-fee interest rate, and $\sigma _ d$ is the standard deviation of the downside:
        $$Sortino_{Ratio} = \frac{R_p - r_f}{\sigma _ d}$$
        \item It is calculated from the time of the funding of the pool and is annualised. The target return in the Sortino Ratio formula is set to 20\%.
        \item The Sortino Ratio calculation has been modified to incorporate a pool's historical relative size. This means that more emphasis is put on performance figures when the total value of the pool is larger vs. smaller.
    \end{itemize}
    \item A pool's Risk Factor is a function of Downside Volatility, which is itself the denominator in the Sortino Ratio calculation. The Risk Factor is 1 for a pool that has had very little Downside Volatility and 5 for a pool with high historical Downside Volatility. To calculate the Score and Risk Factor, dHEDGE requires a minimum management history of 28 days.  
\end{itemize}

dHEDGE aims to connect investors with peers who then mirror their strategy. Investors are protected from withdrawl by the manager because they retain custody of their funds throughout the duration of their pool investment. When an investor buys a stake in a trader's pool, the investor is issued a pool token as opposed to a standard cryptocurrency. The pool token is a derivative token mechanism by which to redeem funds from the smart contracts. Only the pool token holder has access to the funds in the smart contract associated with that user's funds.

\section{Data-Driven Studies}
In this section we discuss various data metrics useful in the analysis of DeFi protocols.

\subsection{Performance Metrics}
Data for these may be found at \url{https://www.defipulse.com/}. Basic metrics for DeFi protocols include:

\begin{itemize}
    \item Total Value Locked (TVL)
    \item Assets Under Management (AUM)
    \item Collateralization Ratio
    \item Trading Volume (DEX, derivatives trading)
    \item Profits
\end{itemize}

\noindent Additionally, various user engagement metrics can be used:
\begin{itemize}
    \item Active users (Daily: DAU,  Monthly: MAU)
    \item New users
    \item Active traders
    \item Active stakers / farmers
    \item Active token holders (governance, voting, delegator)
    \item Active investors (for portfolio management protocols)
\end{itemize}

A Dune analytics database for these metrics can be found at \url{https://dune.xyz/jefftshaw/Balancer}. Metrics relating to risk management, including collarteral ratios and liquidations can be found at \url{https://dune.xyz/bob40/dydx-liquitations}. Additionally, the authors have created their own liquidation dashboard, highlighted in the following section (see Section \ref{dashboard}).

\subsection{Liquidation Dashboard} \label{dashboard}
To get a sense of how liquidation changed as a function of time across major protocols, the team developed a Dune Analytics dashboard analyzing liquidations on two different DeFi protocols: MakerDAO and dYdX. See \url{https://dune.com/matthewding/DyDx-Liquidation} for full data analysis and data tables, including liquidation histories, liquidation event blocks, ETH price, and total users over time.

\begin{figure}[H]
    \centering
    \includegraphics[width=15cm]{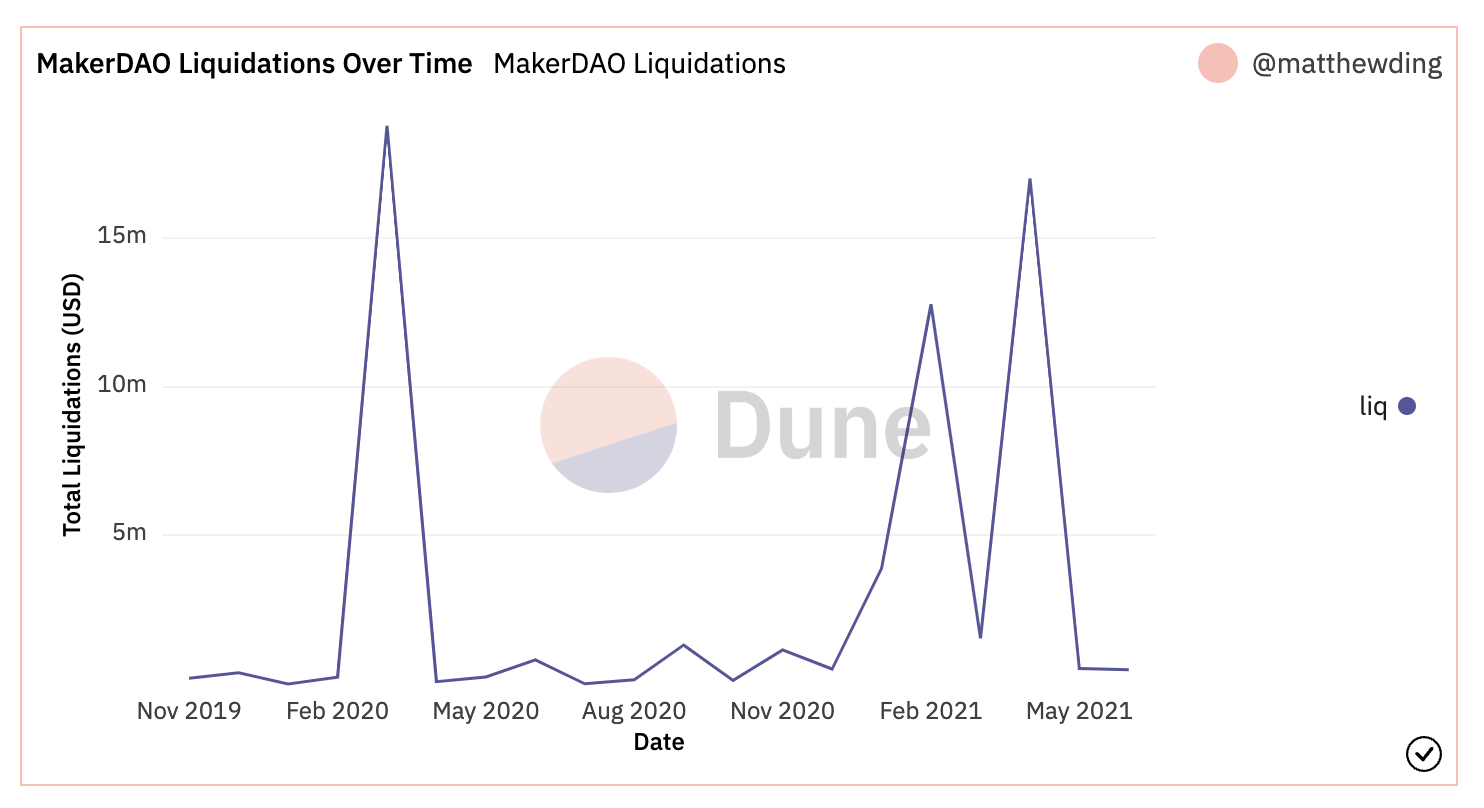}
    \caption{MakerDAO Liquidation History}
    \label{makerdao_history}
\end{figure}

\begin{figure}[H]
    \centering
    \includegraphics[width=15cm]{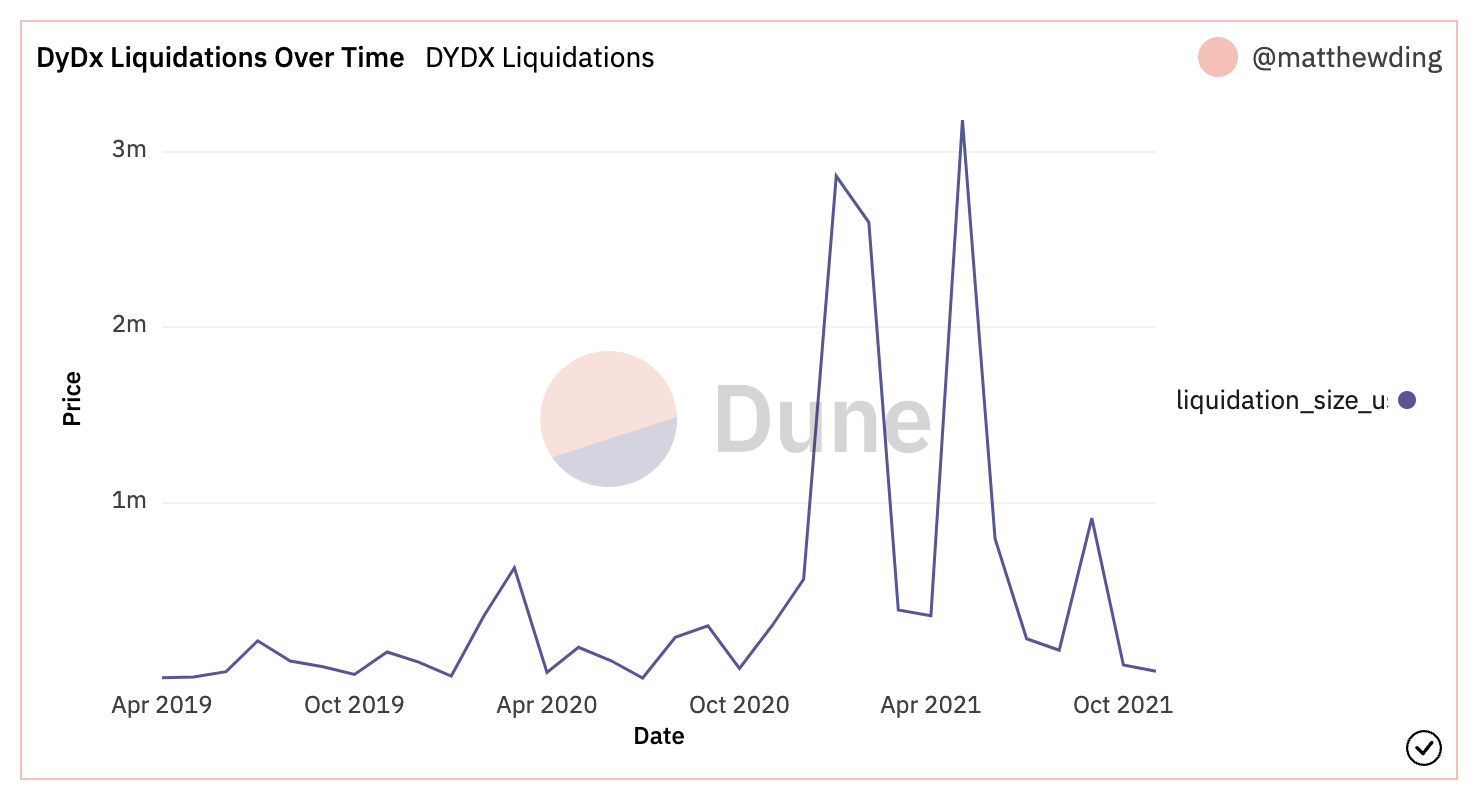}
    \caption{dYdX Liquidation History}
    \label{dydx_history}
\end{figure}

In Figures \ref{makerdao_history} and \ref{dydx_history}, we graph the history of liquidation amounts in (USD) over the lifespan of the protocol. The graphs clearly show major liquidation events in the relative peaks (e.g. March 12, 2020 on MakerDAO [\citeonline{makerdao_liquidation}]).

\begin{figure}[H]
    \centering
    \includegraphics[width=15cm]{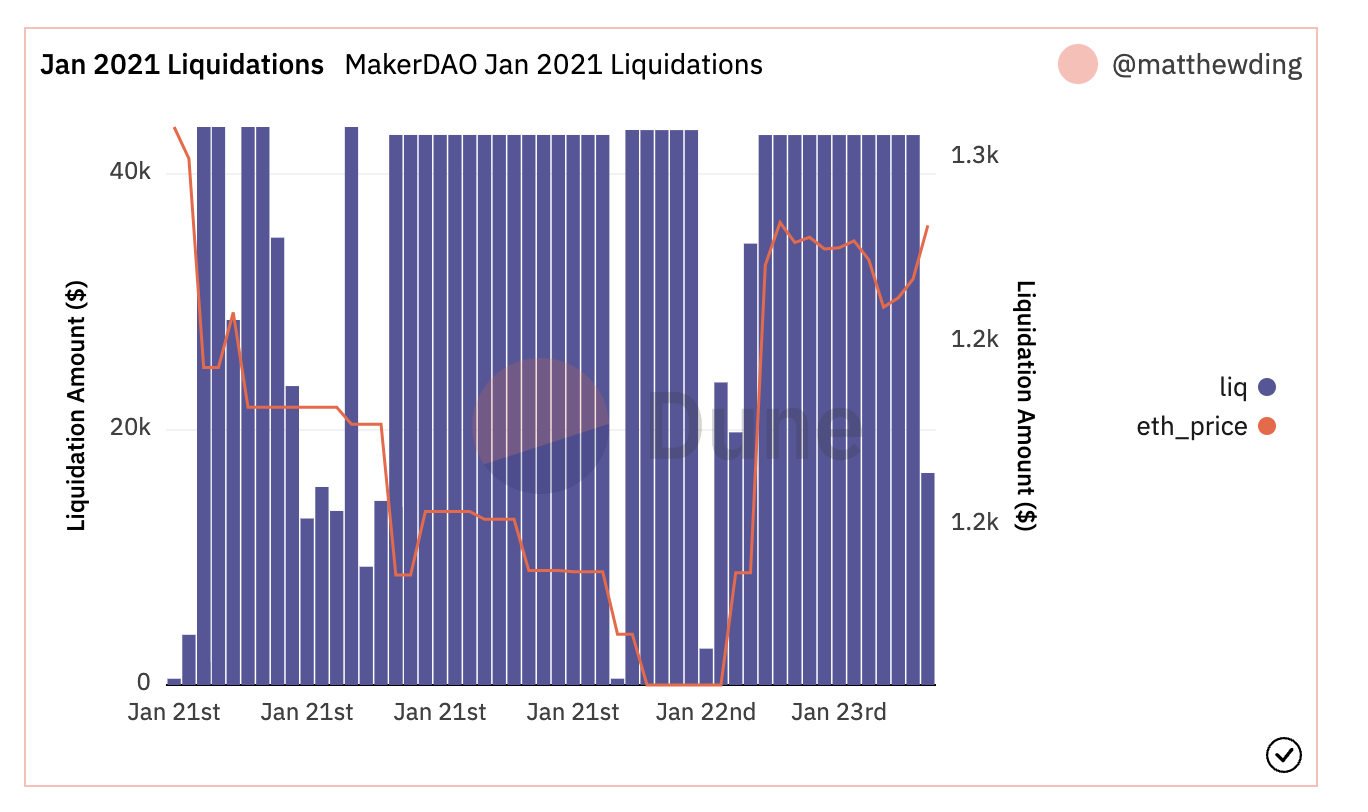}
    \caption{MakerDAO Jan 2021 Liquidations and ETH Price}
    \label{makerdao_liquidation}
\end{figure}

\begin{figure}[H]
    \centering
    \includegraphics[width=15cm]{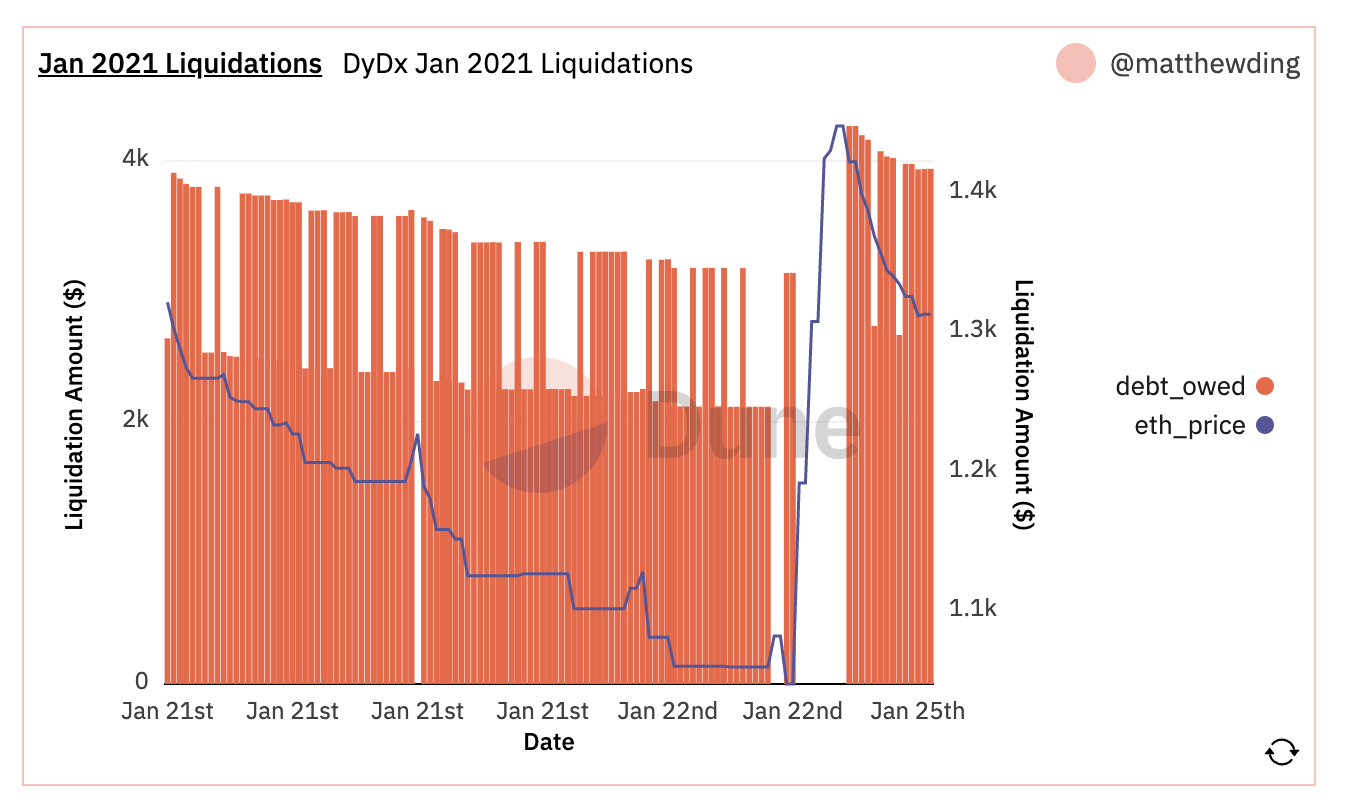}
    \caption{dYdX Jan 2021 Liquidations and ETH Price}
    \label{dydx_liquidation}
\end{figure}

In Figures \ref{makerdao_liquidation} and \ref{dydx_liquidation}, we specifically look at a 5-day long period during January 2021. In these graphs we clearly see an inverse correlation between \textbf{change} in ETH price and the amounts and sizes of liquidation events on these two DeFi protocols. As price decreases, liquidations remain high. While the price spikes on Jan 22, liquidations plummet to low levels.

\section{Formulation of the Key Mechanisms}
\subsection{Synthetic Assets: Basic Mechanisms}

To build the synthetic assets minting/generating system, we should always consider the following mechanisms:
\begin{itemize}
    \item Fee mechanisms: AMM for synthetic assets (Mirror) or unlimited liquidity without AMM (Synthetix)
    \item Liquidation threshold: Using a collateral ratio to manage the synthetic assets generation and destruction (liquidation).
    \item Oracle/Synth pegging mechanism: Using solid oracle services to reduce tracking error, and reduce collisions in the markets. 
    \item Staking: Depositing a cryptocurrency coin or token into a liquidity pool or yield farming protocol.
    \item Neutral debt pool: Used for pooling all the synthetic assets in the investment for all the users, and manages the risk and revenue.
    \item Governance risk: In traditional companies, activist investors can buy shares and vote to tilt the company’s direction as they desire. DeFi protocols with governance tokens are similar, except governance systems are launched at the very inception of a protocol’s life cycle, which can create greater vulnerabilities for a hostile takeover and/or drastic changes to the policies of a protocol.
    \item Scaling solution: For a block to become part of a blockchain, every miner must execute all the included transactions on their machine. To expect each miner to record each and every financial transaction for a global asset market is unsustainable. Potential solutions being implemented include proof-of-stake and Layer 2 chains (ex: optimistic roll-ups). 

\end{itemize}

\subsection{Price Divergence Reduction Mechanisms}
\begin{itemize}
    \item Liquidity incentives on the DEX: Liquidity incentives for the synthetic assets (Mirror protocol), and the portfolio management protocol (tokenSet).
    \item Neutral debt pool: Pooling all the synthetic assets with huge liquidity, to reduce the slippage.
    \item Oracles: Track the real-time price of the pegged/collateralized assets and port all the data on-chain.
    \item Oracle risk: Many DeFi protocols require access to secure, tamper-resistant asset prices to ensure that routine actions such as liquidations and prediction market resolutions function correctly. Protocol reliance on external data feeds introduces oracle risk. An ``oracle attack" is when malevolent actors target price oracles, manufacturing favorable on-chain exchange rates that allow for arbitrage.
    \item Slippage: Just as in TradFi, slippage refers to a difference in price between buyer and seller expectations between the time a transaction is requested versus when it executes. Many protocols seek to limit this.
    \item Token Issuance/Redemption: The portfolio management protocols like Token Set use the token issuance/redemption mechanism, to reduce the slippage from the direct swap of the tokens in the baskets.
\end{itemize}

\subsection{Risk Management Strategies}

\begin{itemize}
    \item Collateralization ratio: The collateralized ratio is also a protection of the healthy ratio between the collateralized assets and synthetic assets. With a relatively high collateralized ratio, the synthetic assets are less likely to be liquidated even under extreme market situations.
    \item Insurance fund mechanism: By using the insurance fund, the protocols are becoming more resilient to sudden market crashes under extreme circumstances. If the insurance fund is depleted, some protocols can have other mechanisms to prevent further loss of the protocols.
    \item Impermanent loss: in most AMM systems, liquidity providers contribute assets to ensure liquidity for traders. If the price of your assets changes drops with respect to when you deposited them, you are exposed to this issue. This dollar value shortfall is known as impermanent loss.
    \item Portfolio management: The on-chain portfolio management protocols allow the users to invest in a basket of the tokens, which reduce the risks or failures to invest in a single token, and can also catch up the trend of certain type of the tokens (e.g., Defi, Metaverse).
    \item Leverage: DeFi is characterised by the high leverage that can be sourced from lending and trading platforms. Funds borrowed in one instance can be re-used to serve as collateral in other transactions, allowing investors to build increasingly large exposure for a given amount of collateral. Derivatives trading on DEXs intrinsically involves leverage, as options payments take place only in the future. The maximum permitted margin in DEXs can be higher than in regulated exchanges in the traditional financial system.
\end{itemize}

\subsection{Decentralized Clearance Mechanisms}
\begin{itemize}
    \item Decentralized liquidation: All the liquidations happen on-chain through executions of the smart contracts. Each liquidator has equal opportunities to participate in the liquidation events of the derivatives.
    \item Liquidity Pools: A reserve of deposited funds meant to provide liquidity to a synthetic asset, blockchain, and/or smart contract. There are typically built-in incentives (ex: mining rewards, transaction fees) to incentivize liquidity providers.
    \item Order matching: The traders on the decentralized derivatives trading platform are matched based on the information of their orders. The protocols (DYDX, etc) seek to find optimal order matching algorithms.
\end{itemize}

\section{Conclusion and Future Work}

Synthetic assets and derivatives protocols are a fundamental building block for the DeFi space, a key asset category emerging as a natural consequence of the growing maturity of technologies and user demands at the intersection of cryptocurrencies and traditional financial products. This systematization of knowledge has put a magnifying glass to the diverse range of DeFi protocols which build, supply data to, trade, or otherwise incorporate crypto derivatives. 

We begin with a wide perspective on the crypto space, the ideological underpinnings thereof, and the regulatory backlash and market uncertainties it is presented with. We establish a general framework for understanding the construction and utility of synthetic assets and apply this understanding towards demystifying the underlying mechanisms, principles, and risks of leading financial tools. We highlight three high-profile synthetic asset protocols: Synthetix, Mirror, and Anchor Protocol. We further comment on the most prevalent derivatives trading platforms in the status quo: dYdX, Deri Protocol, Lyra, Vega, and Thales. This broad scope entailed a focus on the mechanisms, incentives, risks, and legality of these protocols as opposed to a more technical deep dive. 

Future research into derivatives protocols can build upon this work, incorporating new developments and challenges to the space as they emerge.

\section{Acknowledgements}

This work was done under and supported by the Center for Responsible, Decentralized Intelligence—an interdisciplinary research hub affiliated with UC Berkeley Engineering, Blockchain @ Berkeley, the Haas School of Business, Berkeley Law, the Sutardja Center for Entrepreneurship and Technology, Berkeley Data Science, and the Simons Institute for the Theory of Computing. All opinions, findings, and conclusions, or recommendations expressed in this material are those of the authors and do not necessarily reflect the views of our partner organizations.

We would like to give a special thanks to Dawn Song, Christine Parlour, and Yongzheng Jia for their guidance throughout this project.

\bibliographystyle{ieeetr}
\bibliography{citation} 

\end{document}